\documentclass[onecolumn,final]{elsart3p}

% ...
 \usepackage{graphicx}

\usepackage{amssymb}
\usepackage{here}       
\usepackage{rotating}

\newcommand{\refeq}[1]{Eq.~(\ref{eq:#1})}
\newcommand{\reffig}[1]{Fig.~\ref{fig:#1}}
\newcommand{\refsec}[1]{Sec.~\ref{sec:#1}}

\newcommand{\reftab}[1]{Table~\ref{tab:#1}}

\newcommand{\Deg}{^{\circ}}
\newcommand{\Sem}{S_{\rm EM}}
\newcommand{\gcmsq}{\mbox{g}/\mbox{cm}^2}

\newcommand{\Smu}{S_{\mu}}
\newcommand{\Xmax}{X_{\rm max}}
\newcommand{\Xgr}{X_{\rm gr}}
\newcommand{\Xmaxavg}{\langle X_{\rm max} \rangle}
\newcommand{\Smax}{S_{\rm max}}
\newcommand{\DXmax}{DX_{\rm peak}}

\newcommand{\Nmu}{N_{\mu}}
\newcommand{\Nmufit}{N_{\mu\:\rm fit}}

\newcommand{\Nmuavg}{\langle N_{\mu} \rangle}
\newcommand{\sigmaNmu}{\sigma_{\Nmu}}

\renewcommand{\v}[1]{\mathbf{#1}}
\newcommand{\vphat}{\v{\hat{p}}}
\newcommand{\vghat}{\v{\hat{g}}}
\newcommand{\vahat}{\v{\hat{a}}}

\begin{document}

\begin{frontmatter}



\title{A Model-Independent Method of Determining Energy
Scale and Muon Number in Cosmic Ray Surface Detectors}


\author{Fabian Schmidt}, \author{Maximo Ave}, \author{Lorenzo Cazon}
\address{Department of Astronomy \& Astrophysics, The University of Chicago,
Chicago, IL~~60637-1433}
\address{Kavli Institute for Cosmological Physics,
Chicago, IL~~60637-1433}
\author{Aaron Chou}
\address{Center for Cosmology and Particle Physics, New York University, 
4 Washington Place, New York, NY~~10003}
\address{Fermi National Accelerator Laboratory, PO Box 500, Batavia, IL~~60510}

\begin{abstract}
Surface detector arrays are designed to measure the spectrum and composition
of high-energy cosmic rays by detecting the secondary particle flux of
the Extensive Air Showers (EAS) induced by the primary cosmic rays.
Electromagnetic particles and muons constitute the dominant contribution 
to the ground detector signals. In this paper, we show
that the ground signal deposit of an EAS can be described in terms of only very
few parameters: the primary energy $E$, the zenith angle $\theta$,
the distance of the shower maximum $\Xmax$ to the ground, and a muon flux 
normalization $\Nmu$. This set of physical parameters is sufficient 
to predict the average particle fluxes at ground level to around 10\% accuracy. 
We show that this is  valid for hadronic air showers, using the two 
standard hadronic  interaction models used in cosmic ray physics, QGSJetII 
and Sibyll, and
for hadronic primaries from protons to iron. 
Based on this model, a new approach to calibrating the energy scale of ground
array experiments is developed, which factors out the 
model dependence inherent in such calibrations up to now.
Additionally, the method yields a measurement of the average number of 
muons in EAS. The measured distribution of $\Nmu$ of cosmic ray
air showers can then be analysed, in conjunction with measurements of $\Xmax$ 
from fluorescence detectors, to put constraints on the cosmic ray composition
and hadronic interaction models.
\end{abstract}

\end{frontmatter}

\section{Introduction}
\label{sec:intro}

The origin of Ultra-High Energy Cosmic Rays (UHECR, with 
energies~$E>10^{18}$~eV) still remains a mystery. 
Experimental results \cite{HPphoton,photonreview,Augerphoton} suggest that 
the UHECR flux is composed predominantly of hadronic primary particles.
As charged particles, they suffer deflections in cosmic magnetic fields and do 
not point
back directly to their sources. Indirect proofs of their origin are necessary
instead: the precise measurement of the energy spectrum, an
estimation of the mass composition and its evolution with energy, and 
angular anisotropies are the three main handles on disentangling
this almost century-old problem. Due to the interaction with the cosmic
microwave background, UHECR suffer energy losses which limit their 
propagation distance \cite{GZK1,GZK2}. This ``GZK horizon'', indications
of which have already been observed in the UHECR spectrum 
\cite{hiresGZK,augericrcSpectrum,AugerScience}, depends 
very sensitively on the energy and mass of the cosmic ray. 
In order to discern between different source scenarios, and to disentangle source
characteristics from the effects of propagation, a precise knowledge of the 
energies of UHECR is crucial. Constraints on the composition of the cosmic ray
flux at the highest energies will supply additional fundamental insight.

Due to the low fluxes at ultra-high energies, the detection of UHECR can
only be achieved by measuring {\it Extensive Air Showers} (EAS),
cascades of secondary particles
resulting from the interaction of the primary cosmic rays with the Earth's 
atmosphere. The measurement of the cosmic ray energy, flux, and mass 
composition relies on an understanding of this phenomenon.

Two main EAS detection techniques were developed over the years
(see \cite{NaganoWatson} for a review):
{\it surface detectors} (SD) detect the particle flux of an EAS at a
particular stage of the shower development; {\it fluorescence
detectors} (FD) measure the shower development through nitrogen
fluorescence emission induced by the electrons in the shower. 
 The modeling of EAS through Monte Carlo
 simulations is needed in both fluorescence and surface detector experiments 
 in order to interpret the
 data. We will show that hadronic EAS can
 be characterized, to a remarkable degree of precision, by only three
 parameters: the primary energy $E$, the depth of shower maximum
 $\Xmax$, and an overall normalization of the muon component, which we
 call $\Nmu$. This is what we will call {\it air shower universality}
 \cite{univicrc}.
 The parameters $\Xmax$ and $\Nmu$ are linked to the mass of the
 primary particle, ranging from proton to iron, and are subject to
 shower-to-shower fluctuations; proton showers have a larger depth of
 shower maximum than iron showers, while iron showers contain $\sim 40$\%
 more muons than those induced by protons. 
 Once measured, $\Nmu$ and $\Xmax$ have to be compared with simulations
 to infer the cosmic ray composition and place constraints on hadronic
 interaction models. Previous studies have demonstrated that the
 energy spectra and angular distributions of electromagnetic particles
 \cite{Nerling,Giller}, as well as the lateral distribution of energy
 deposit close to the shower core \cite{Gora} are all universal, i.e.
 they are functions of $E$, $\Xmax$, and the atmospheric depth $X$
 only.\footnote{The dependence on $X$ and $\Xmax$ is commonly put in
   terms of the shower age $s$. We will use a different parameter, $DX$, which
is better suited to our purpose.} For studies of shower universality in the
context of ground detectors, see \cite{billoirphoton,GAP1,GAP2}.
EAS induced by photons show somewhat different properties, due to the
absent hadronic cascade. Hence, it remains to be investigated to what
extent the hadronic EAS universality studied here applies to photon
showers.

By sampling the longitudinal development of the
electromagnetic shower component close to the core, fluorescence detectors
measure both $\Xmax$ and $E$.
The systematic uncertainty in the energy $E$ is typically 25\%,
mainly due to the uncertainties in the air fluorescence yield.
A surface detector only samples the properties of
an EAS at a given stage of the shower development and
at several points at different distances $r$ from the shower
axis. Rather than using the signal integrated over all distances, a quantity
which shows large fluctuations, Hillas \cite{Hillas} proposed to use the 
signal at a given distance $r$ from the shower axis, $S(r)$, 
as a measure of the shower {\it size}, 
connected with the primary energy. The distance where
experimental uncertainties in the {\it size} determination are minimized
(the optimal distance $r_{\rm opt}$ \cite{Newton}) is mainly
determined by the experiment geometry, i.e. the spacing between surface
detectors.  $S(r_{\rm opt})$ is then related with the primary energy
of the incoming cosmic ray using Monte Carlo simulations. This
calibration has large systematics due to uncertainties in the
hadronic models and the unknown
primary cosmic ray composition.

In this paper, we will show how to use air shower universality
to determine the calibration of a surface detector in a 
model-independent way. The signal $S(r_{\rm opt})$ is the sum of two
components: an electromagnetic part which is well-understood and to a
good approximation depends only
on $E$ and $\Xmax$ of the shower; and a muon part which, in addition to
$E$ and $\Xmax$, depends on the model and primary composition in terms of
an overall normalization.
The muon fraction can be determined by requiring that the
shape of the zenith angle dependence of $S(r_{\rm opt})$ at a
fixed energy, which depends on the muon normalization $\Nmu$,
match the observed one. This method determines the energy scale of the 
experiment as well as the
average number of muons produced in the air showers at a given energy. 

Subsequently, we will apply air shower universality to data
collected by a {\it hybrid experiment}, which combines the fluorescence
technique with a surface detector. In this case, the
calibration of the surface detector can be done almost independently
of hadronic models and composition by using a small subset
of the data (hybrid events) which are simultaneously measured by
the fluorescence and the surface detector. Applying our method to
hybrid data yields an event-by-event measurement of the muon content
of the shower. This can be used as an independent cross-check of the
measurement from the surface detector alone. Since the electromagnetic
contribution to the signal varies with zenith angle, a hybrid measurement 
of $\Nmu$ at different zenith angles probes whether the electromagnetic part is
described correctly by simulations, a key ingredient in our study.
Conversely, the surface detector energy scale obtained with the
universality-based method offers a cross-check of the hybrid calibration 
of the surface detector, which uses the fluorescence energy measurement.

In this work, we will not use data from any experiment, but we will use the
Pierre Auger Observatory as a case of study. First results from this method
applied to Auger data have already been presented in \cite{augericrcNmu}. 
While we adopt the specifications of this experiment, the method presented
here can be applied to any other surface detector (for example, 
AGASA \cite{agasa})
or hybrid experiment (for example, Telescope Array \cite{TA}).
The paper is organized as follows: in \refsec{univ} we explain 
air shower universality, verifying it with the two standard high energy 
hadronic models used in cosmic ray physics (QGSJetII and Sibyll);
in \refsec{viol} the limits of air shower universality are shown; 
\refsec{CIC} presents the method of obtaining $\Nmu$ from the surface 
detector and determining the surface detector energy scale; in \refsec{toyMC} 
we validate the method using a simple
Monte Carlo approach; \refsec{hybrid} shows how the approach can be applied
to hybrid events; finally, the application to other experiments is
discussed in \refsec{otherexp}; we conclude in \refsec{disc}.

\section{Extensive air shower universality at large core distances}
\label{sec:univ}

The results presented in this paper were obtained from a library of
simulated EAS. We used CORSIKA 6.500 \cite{corsika} with hadronic interaction 
models QGSJetII \cite{qgsjet,qgsjet2} and Fluka \cite{fluka} (proton and 
iron primaries, energies
$10^{17.8}-10^{20}$~eV), and Sibyll \cite{sibyll,sibyll2} / Fluka as well as
QGSJetII / Gheisha \cite{gheisha} 
(proton at $10^{19}$~eV). For each primary/energy
combination, we simulated 80 showers each at 7 zenith angles ranging
from $0\Deg$ to 60$\Deg$. Statistical thinning was employed in the simulations
as describeds in \cite{Kobal}, at a thinning threshold of 
$\varepsilon=10^{-6}$.

Using lookup tables generated with GEANT4
simulations \cite{geant4}, we calculated the average response of a cylindrical 
water Cherenkov detector
(height 1.2m, cross section 10~m$^2$, similar to the type used in the
Pierre Auger observatory) to each shower particle hitting the ground.
See \refsec{otherexp} for a discussion of the applicability to other
experiments.
The signals were calculated in two different approaches: 1.)~{\it
  Ground plane signals:} The response is calculated for a realistic
water tank on the ground. 2.)~{\it Shower plane signals:} The 
response is calculated for a fiducial flat detector (with the same average 
particle response as the water tank) placed in the plane
orthogonal to the shower axis ({\it shower plane}). Signals
calculated in the shower plane procedure are not affected by detector
geometrical effects, and therefore independent of the zenith angle.  For
details on the signal calculation, see appendix
\ref{sec:SP}. The shower plane signals will be useful to verify 
air shower universality, while the ground plane signals
will be needed for the application to a realistic experiment (in our case the 
Pierre Auger Observatory).

Due to the statistical thinning procedure 
employed in the shower simulation, 
particles were collected in a sampling area of width 0.1 in $\log_{10}\:r$ 
centered around the shower
core distance $r$ considered. This ensures, for a wide range of slopes of the
lateral distribution, that the median radius of the 
energy deposit in the sampling area is indeed $r$. 
Signals are calculated in 18 azimuthal sectors,
and normalized relative to the signal deposited by a vertically incident
muon (VEM), a standard practice in surface detectors using the water
Cherenkov technique.

\begin{figure}[t]
\begin{minipage}[t]{0.48\textwidth}
\center
\includegraphics[width=\textwidth]{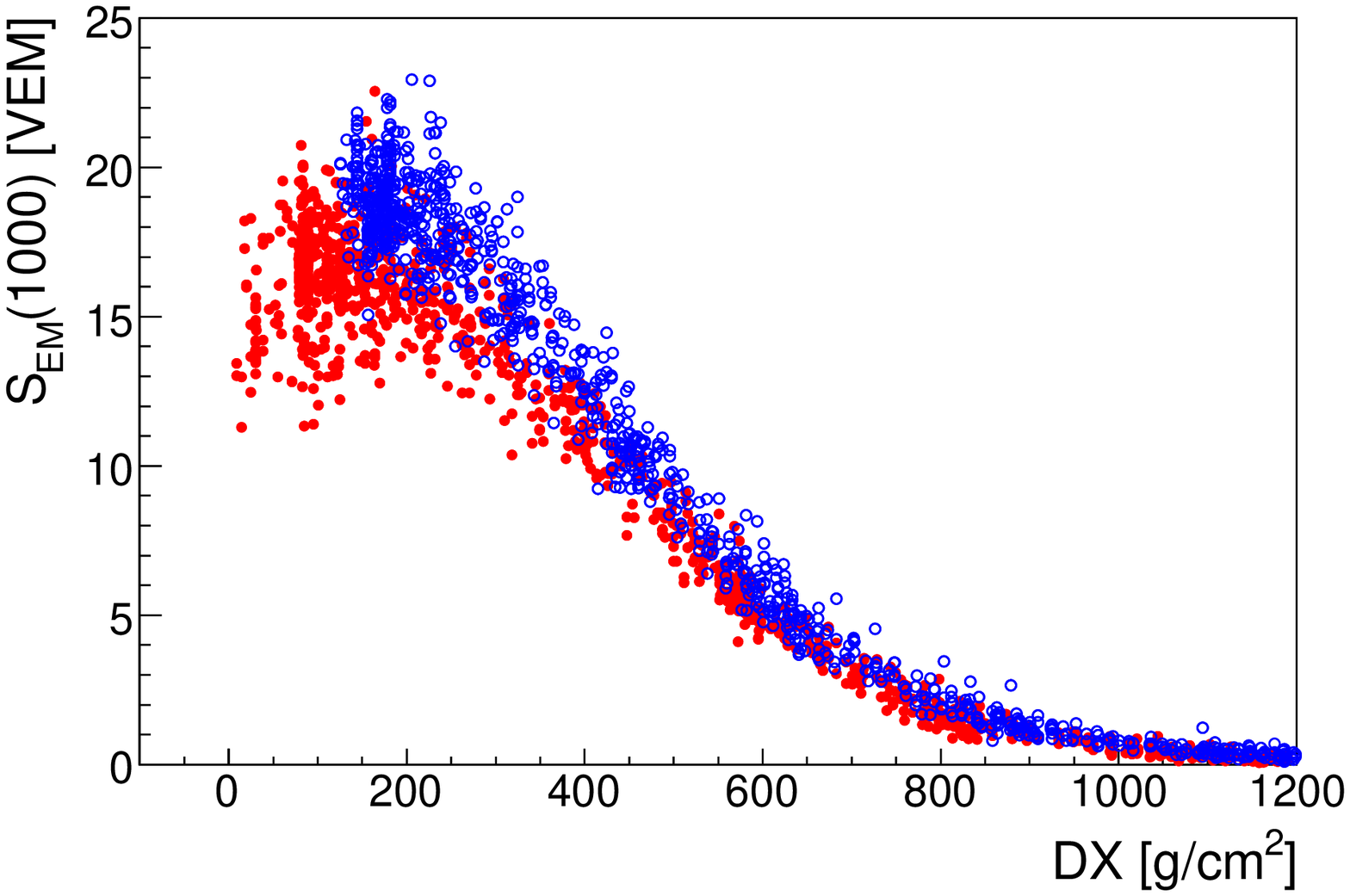}
\end{minipage}
\hfill
\begin{minipage}[t]{0.48\textwidth}
\center
\includegraphics[width=\textwidth]{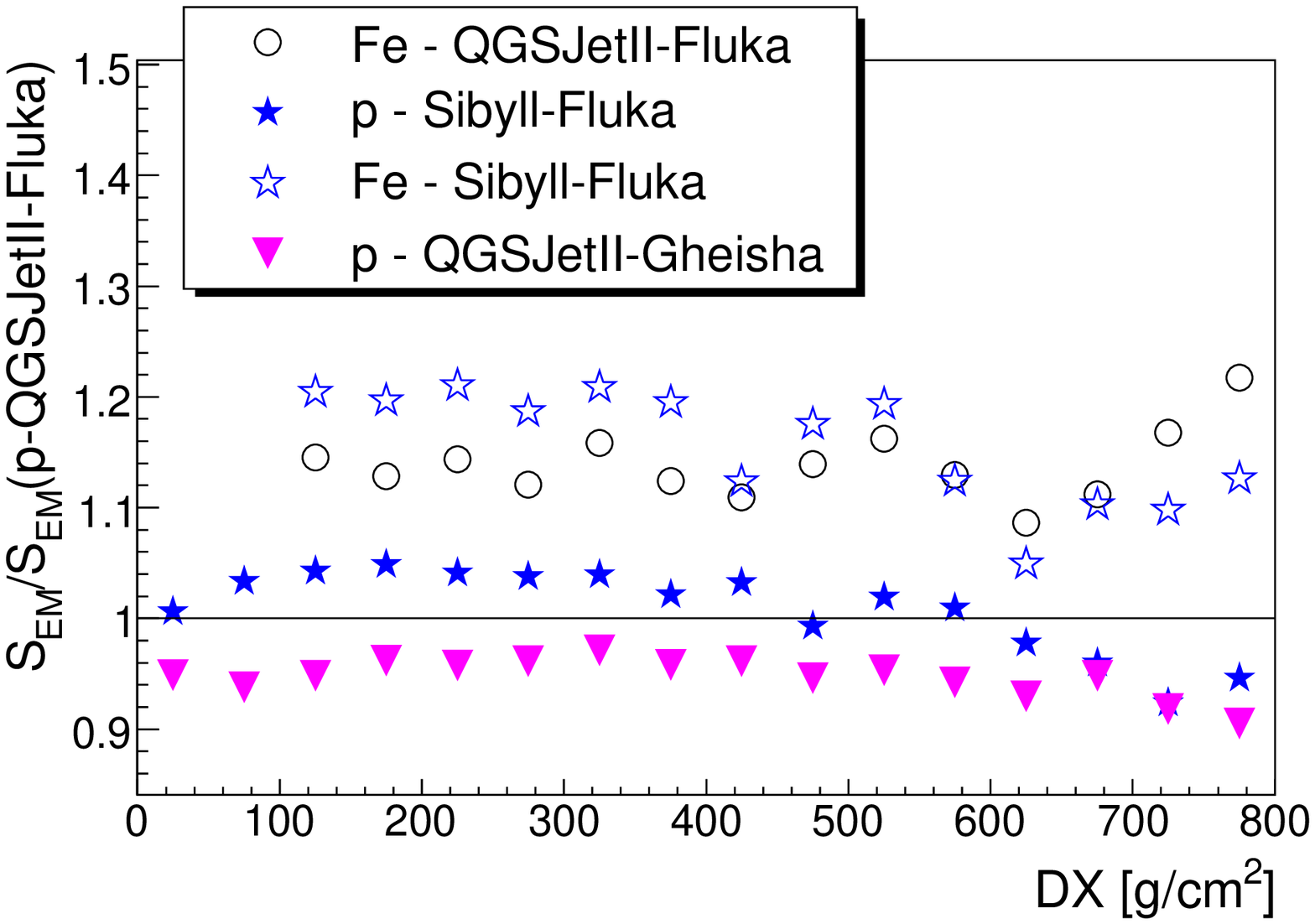}
\end{minipage}
\caption{{\it Left panel:} Simulated electromagnetic shower plane signals 
at $r=1000$~m for proton (red dots) and iron showers (blue circles) at 
$10^{19}$~eV as a function of $DX$.
The showers are simulated with QGSJetII/Fluka at discrete zenith angles 
spanning 0$\Deg$ to $60\Deg$. {\it Right panel:} Simulated electromagnetic 
shower plane signals vs. $DX$ for different primaries and hadronic  
models, relative to the prediction for proton showers when using QGSJetII/Fluka.}
\label{fig:Sem}
\end{figure}

To describe the stage of the shower development,
we use the variable $DX$, defined as the distance from the detector to the shower 
maximum measured along the shower axis (in $\gcmsq$). For a tank on ground at 
a distance $r$ from the shower axis, $DX$ is:
\begin{equation}
DX = \Xgr\:\sec\theta -\Xmax - r\:\cos\zeta\:\tan\theta\:\rho_{\rm air}
\label{eq:DX}
\end{equation}
where $\Xgr$ is the vertical depth of the atmosphere, $\theta$ is the zenith 
angle, $\Xmax$ is the slant depth of shower maximum, and $\zeta$ is the
azimuthal angle in the shower plane such that $\zeta=0$ corresponds to 
a tank below the shower axis.  
$\rho_{\rm air}\approx 10^{-3}\mbox{g}/\rm cm^3$ is the density
of air at ground level (see also \reffig{asym_sketch} in the appendix). 
Often, we will consider signals averaged over azimuth. In this case,
$DX$ is simply given by:
\begin{equation}
DX = \Xgr\:\sec\theta - \Xmax
\label{eq:DX-Xmax}
\end{equation}

\subsection{Electromagnetic and muon shower plane signals}
\label{sec:signals}

At large core distances ($r \gtrsim 100$~m), the particle flux of EAS at ground
is dominated by electromagnetic particles ($e^+$, $e^-$, $\gamma$) and
muons. Throughout the paper, we include the signal from the electromagnetic 
products of in-flight muon decay in the muon contribution, separating it from 
the `pure' electromagnetic part. 

\reffig{Sem} (left panel) shows the electromagnetic shower plane signals $\Sem$
of simulated proton and iron showers at $10^{19}$~eV as a function of
$DX$ (in $\gcmsq$) for a core distance of 1000~m. For each tank, we calculate
the corresponding $DX$ via \refeq{DX} using the azimuth angle of the tank.
Since zenith angle
dependent detector geometry effects are removed in the shower plane
treatment, we are able to compare the signals from a wide range of
zenith angles. The electromagnetic signal shows a strong evolution
with $DX$, reaching a maximum at $\DXmax$ and rapidly
attenuated for larger $DX$.  $\DXmax$ depends
on core distance, being $0\:\gcmsq$ very close to the core and
$\approx 200\:\gcmsq$ at 1000~m. This shift is only mildly dependent
on $r$ in the range $400-1600$~m and it can be naturally explained
by diffusion of electromagnetic particles away from the shower axis.
Note that the overall electromagnetic signal as well as its evolution
are slightly different for protons and iron. This is apparent in the
right panel of \reffig{Sem}, where the ratio of the signals obtained from
different primary/model combinations to proton-QGSJetII is shown as
a function of $DX$. The differences between models are around 5--10\%, smaller 
than the deviation between proton and iron. This result is an extension to 
large $r$ of previous
results \cite{Gora,Giller,Giller2} on the universality of the
electromagnetic EAS component at small core distances.  We address the
difference ($\sim$ 15\%) in $\Sem$ between protons and iron in
\refsec{viol}.

\begin{figure}
\begin{minipage}[t]{0.48\textwidth}
\center
\includegraphics[width=\textwidth]{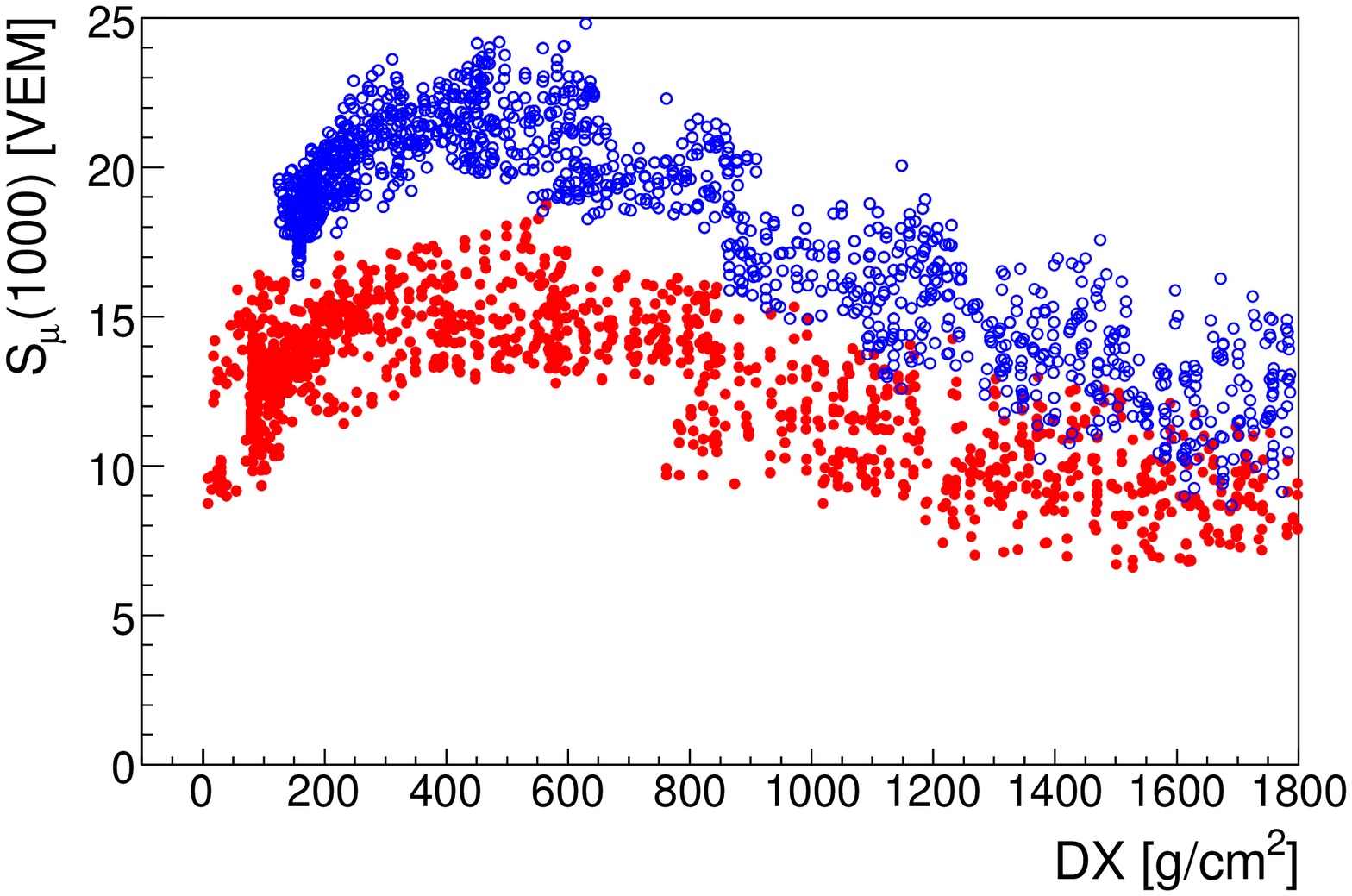}
\end{minipage}
\hfill
\begin{minipage}[t]{0.48\textwidth}
\center
\includegraphics[width=\textwidth]{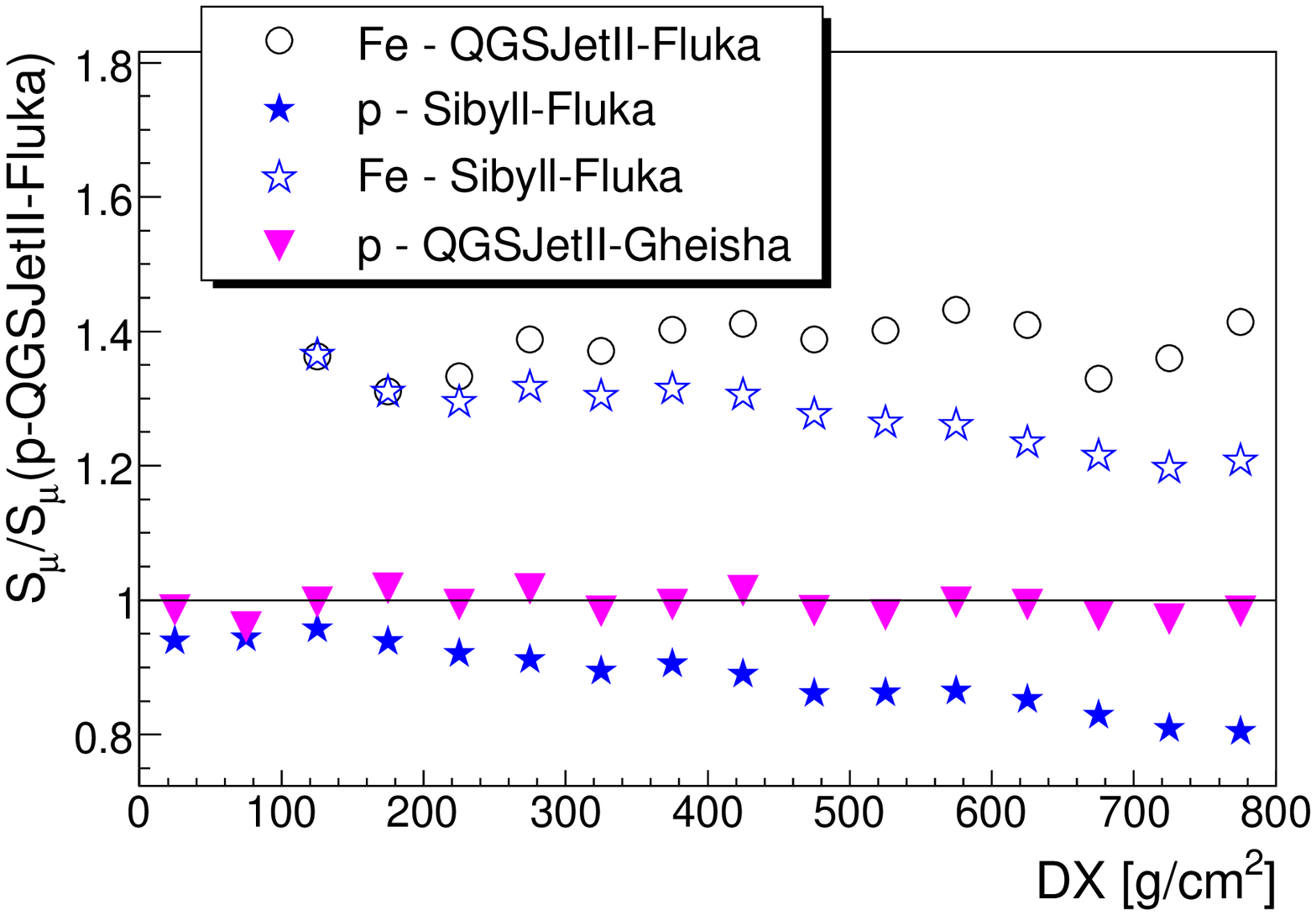}
\end{minipage}
\caption{{\it Left panel:} Simulated muon signals $\Smu$ at $r=1000$~m vs. $DX$ 
for the same $10^{19}$~eV proton (red dots) and iron
showers (blue circles) as in \reffig{Sem}. Note that $\Smu$
includes the contribution from muon decay products.
{\it Right panel:} Simulated muon signals vs. $DX$ for different
primaries and hadronic models relative to the muon signal predicted for
proton showers when using QGSJetII/Fluka. Note the difference in scale compared
to \reffig{Sem}.}
\label{fig:Smu}
\end{figure}

\reffig{Smu} (right panel) shows the evolution of the muon signal $\Smu$ with
$DX$ (again for proton and iron showers at $E=10^{19}$~eV and $r$=1000~m).
$\Smu$ shows a
distinctly different behavior: it peaks at $\DXmax \approx
400\:\gcmsq$, and it is attenuated much more slowly than $\Sem$.
As expected, there is a dependence of the absolute normalization
of the signals on the primary particle and hadronic model, which is clearly
seen in \reffig{Smu} (right panel) where we again show $\Smu$
obtained for different primaries and models relative to that of 
proton-QGSJetII.
As for $\Sem$, only differences in normalization and not in shape are apparent.
We verified that the primary- and model-independence of the electromagnetic
and muon signal evolution holds for shower core distances between 100~m and
1000~m. 

\subsection{Ground signal parameterization}
\label{sec:param}

In the previous section, we have shown that the evolution of the shower plane 
signals at a given shower core distance is only very weakly dependent on the
primary particle or hadronic model considered. Therefore, a simple
parameterization of the signals is possible. In this work, since we use
the Pierre Auger Observatory as a case of study, we perform such a
parameterization for $r$=1000~m. It has been shown \cite{augericrcS1000}
that the main observable in the surface detector of the Pierre Auger
Observatory, $S(1000)$, is indeed a good measurement of the
azimuth-averaged signal of particles at a core distance of $r=1000$~m.
Hence, we separately parameterize the azimuth-averaged ground plane
electromagnetic and muon signal at $10^{19}$~eV ($\Sem(1000)$ and
$\Smu(1000)$, the total predicted signal being the sum of both), using
the incomplete gamma, or Gaisser-Hillas-type function:
\begin{equation}
S(1000,\:DX) = \Smax\left ( \frac{DX-X_0}{\DXmax-X_0} \right )^{\alpha}
\exp \left ( \frac{\DXmax - DX}{\lambda} \right ), \;\; \alpha \equiv \frac{\DXmax-X_0}{\lambda}\;
\label{eq:gh}
\end{equation}
The four free parameters of this function are: $\Smax$ (the peak signal
at 1000~m);
$\DXmax$(the slant depth relative to the overall shower maximum where 
the peak signal is reached); 
$\lambda$ (the attenuation length
after the maximum); and $X_0$ (an additional shape parameter).

\begin{figure}[t]
\begin{minipage}[t]{0.48\textwidth}
\center
\includegraphics[width=\textwidth]{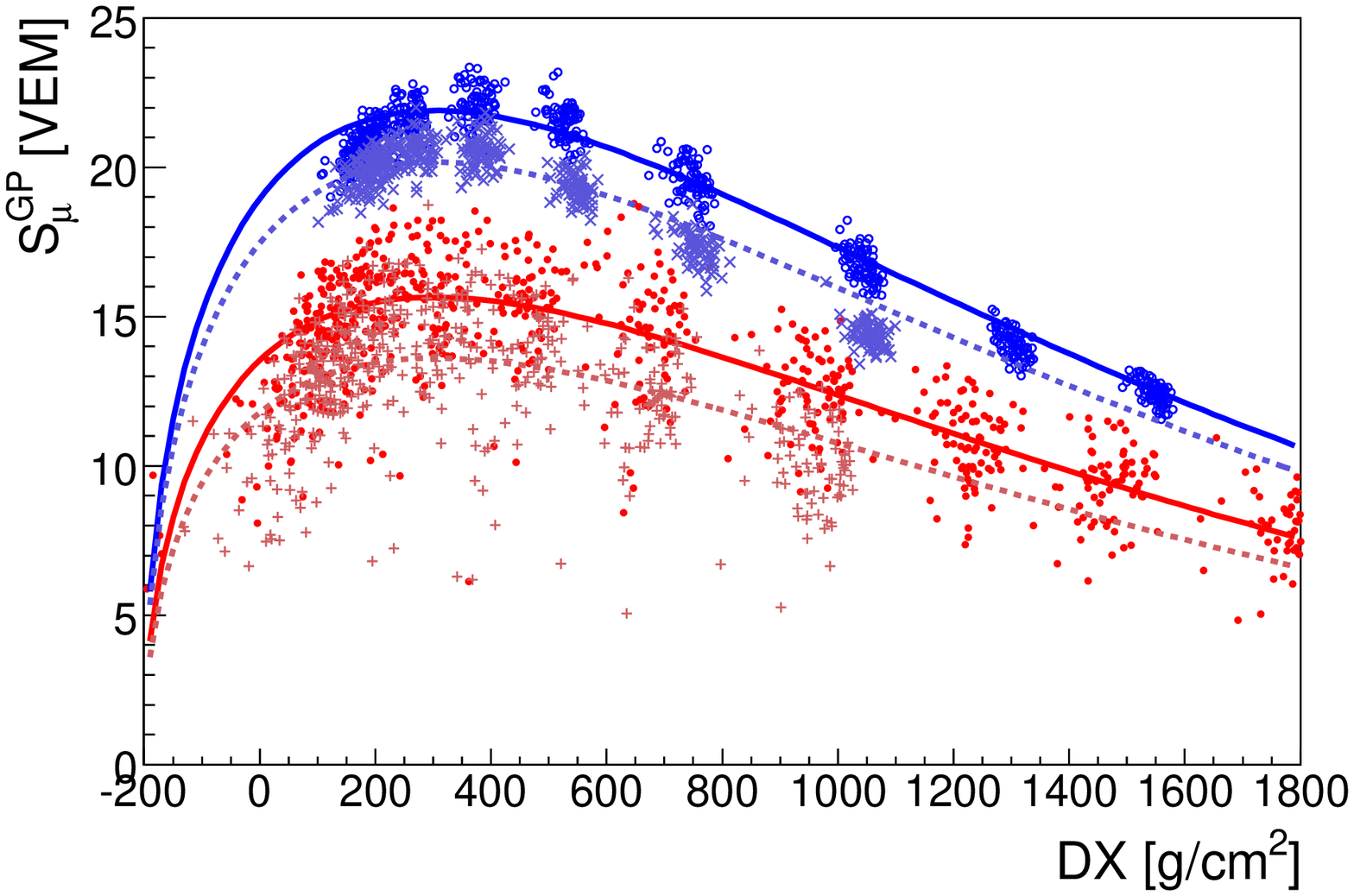}
\caption{Parameterization of the muon ground plane signal ($E=10^{19}$~eV)
at $r=1000$~m using Gaisser-Hillas functions (\refeq{gh}). Red dots (crosses)
denote proton-QGSJetII (proton-Sibyll) showers, blue circles (asterisks) are
iron-QGSJetII (iron-Sibyll).
The normalization
is left free for each model/primary combination, while the other parameters
are common to all.}
\label{fig:Smufit}
\end{minipage}
\hfill
\begin{minipage}[t]{0.48\textwidth}
\center
\includegraphics[width=\textwidth]{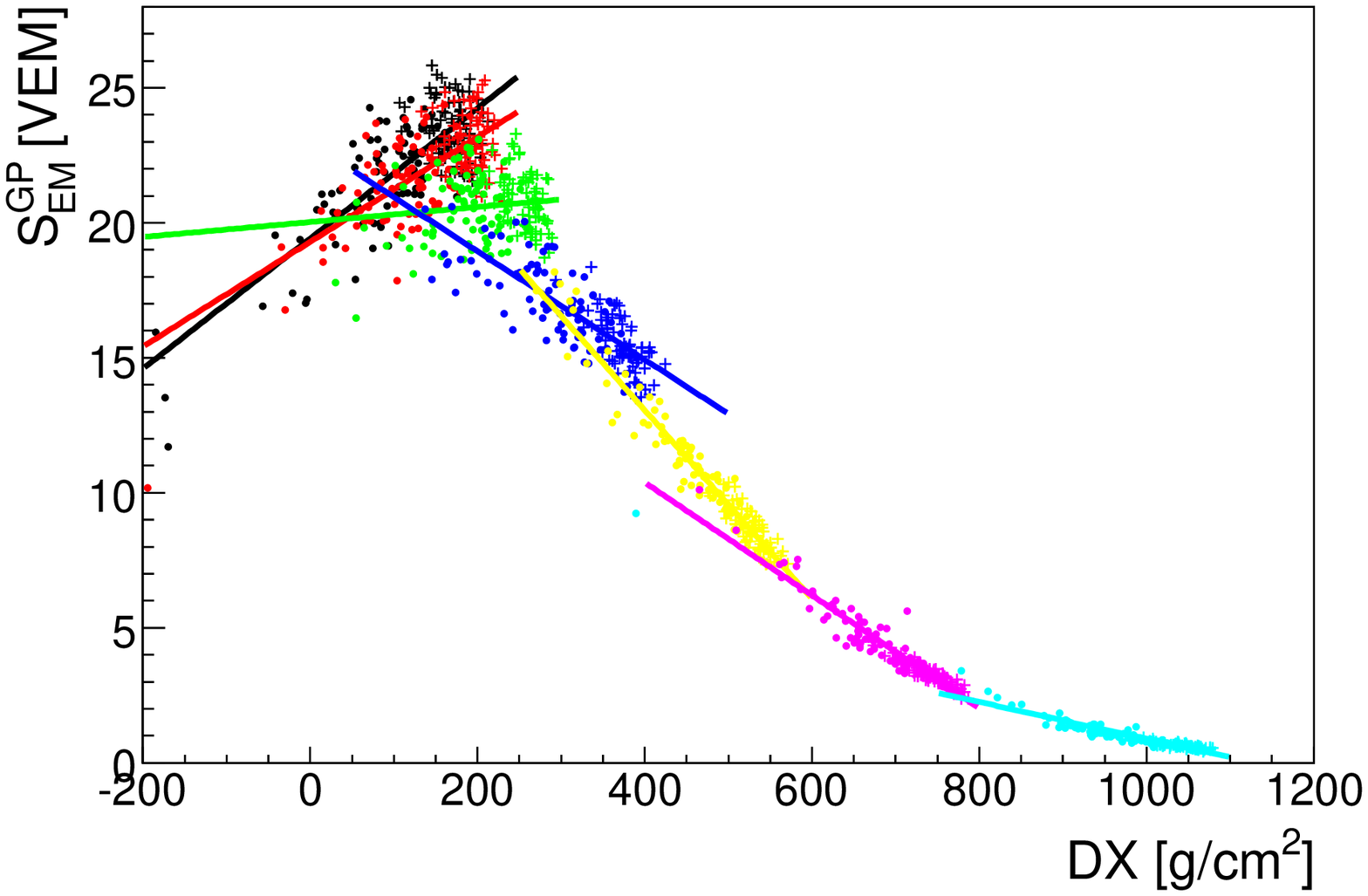}
\caption{Electromagnetic ground plane signals (proton$-$dots; iron$-$crosses;
QGSJetII) in zenith angle bands (color-coded). Proton and iron signals have
been scaled symmetrically. A linear function is fit to
the signal separately for each zenith angle band.}
\label{fig:Semfit}
\end{minipage}
\end{figure}

\reffig{Smufit} shows the results of the fit for the muon signal.
We have simultaneously fitted the predictions for different primaries
(proton, iron) and different models (QGSJetII, Sibyll), keeping a
separate normalization ($\Smax$) for each, while $\lambda$ and
$\DXmax$ are common to all. $X_0$ is not fitted but fixed to
$-200\:\gcmsq$. The resulting parameters are summarized in
\reftab{fitpar}, with $\Smax$ given for proton-QGSJetII. For the other
model/primary combinations, we define a {\it relative} muon
normalization given by $\Nmu = \Smax/S_{\rm max;ref}$, where we take
proton-QGSJetII as the reference $S_{\rm max;ref}$. The $\Nmu$ for
different models and primaries are listed in \reftab{sh2sh}.

In the case of the electromagnetic signal, we have to take into
account the detector geometrical effects, which cause differences in the 
signals from showers at two different zenith angles with the same $DX$
(see appendix~\ref{sec:SP}).  
Hence, we have to find a parameterization for
$\Sem(DX,\:\theta)$. The first step is to parameterize, for each of the 7 
simulated zenith angles, the dependence of $\Sem$ on $DX$; a linear function is
found to be sufficient due to the limited $DX$ range at a fixed
$\theta$. We scaled the proton and iron signals by $1+\alpha$ and $1-\alpha$,
respectively (with $\alpha\lesssim 0.06$), to account for the deviations from 
universality. The deviations 
in the ground plane signals are slightly smaller than those shown in the 
previous section for the shower plane signals.
\reffig{Semfit} shows the results of the fits together with the direct Monte 
Carlo results, for the 7 fixed values of zenith angle. 

In the second step, we fit a Gaisser-Hillas-type function (of $DX$) to 
$\Sem(DX,\theta)$ for each $\Xmax$ considered. The Gaisser-Hillas function is 
fitted to 7 equal-weight data points ``predicted'' from the 7 linear fits of the 
first step. This is equivalent to a parameterization of the dependence of 
$\Sem$ on $\theta$ at a fixed $\Xmax$, using an intermediate variable 
$DX(\Xmax,\theta)$. \reftab{fitpar} gives the results for $\Xmax = 750\:\gcmsq$
and $\Xgr=875\:\gcmsq$.  
The second step may be applied to any value of $\Xmax$, yielding a continuous 
function $\Sem(\theta,DX(\Xmax,\theta))$ which depends on $\theta$ both
explicitly and implicitly via $DX$. By contrast, since muons deposit a signal 
which is proportional to their pathlength in the water tank, the tank acts as 
a volume detector: the smaller projected area at higher zenith angle is canceled
by the longer average tracklength. Hence, the average muon signal $\Smu(DX)$ does not show
an explicit $\theta$ dependence.

\begin{table}[b]
\center
\begin{tabular}{cc|c|c|c|c}
\hline
& & ~$\Smax$~ & ~$\DXmax$~ & ~~~$X_0$~~~ & $\lambda$ \\
\hline
$\Sem(1000)$ & ($\Xmax=750\:\gcmsq$) & 22.5 & 103.0 & -540.6 & 102.7 \\
$\Smu(1000)$ &  & 15.6 & 302.4 & -200 & 1109 \\
\hline
\end{tabular}
\caption{Fit parameters of the Gaisser-Hillas parameterization (\refeq{gh}) 
of the universal  electromagnetic and muon signal at $10^{19}$~eV. The
electromagnetic parameterization is for a fixed $\Xmax$ and $\Xgr=875\:\gcmsq$. 
For $S_{\mu\:\rm max}$,
the value of proton-QGSJetII is given (for the other primaries and models
relative to proton-QGSJetII, see \reftab{sh2sh}).}
\label{tab:fitpar}
\end{table}

At a fixed energy (here, 10~EeV), the parameterization presented above determines the 
average ground signal of a shower (at $r=1000$~m, azimuth-averaged):
\begin{equation}
S(1000) = \Sem(\theta,\:DX(\Xmax,\theta)) + \Nmu\cdot S_{\mu;\rm ref}(DX(\Xmax,\theta))
\label{eq:Sparam}
\end{equation}
Here, $\Sem$ denotes the parameterized electromagnetic signal, and 
$S_{\mu;\rm ref}$ is the reference muon signal which we take to be 
proton-QGSJetII. Hence, there are only three free parameters describing
the average shower at this energy: the zenith 
angle $\theta$; the depth of shower maximum $\Xmax$; and the normalization
of the muon signal $\Nmu$ (relative to proton-QGSJetII).

We used the library of proton and iron showers with energies of
$10^{18}-10^{20}$~eV to investigate the energy dependence of the evolution
of $\Sem$ and $\Smu$ with $DX$. The electromagnetic signal normalization
shows an energy scaling of $S_{\rm max; EM}\propto E^{0.97}$ (see also
\refsec{viol}), while for the muon signal $S_{\rm max; \mu}\propto E^{\alpha}$ 
with $\alpha=0.9\dots 0.95$, depending on the hadronic model. All other
fit parameters in \refeq{gh} are independent of the primary energy in this energy 
range, for both $\Sem$ and $\Smu$, to within 5\%.

Hence, \refeq{Sparam} can be straightforwardly extended to other energies:
\begin{eqnarray}
S(1000,E) &=& \Sem(10\:\mbox{EeV},\theta,\:DX(\Xmax,\theta))\: \left ( \frac{E}{10\:\mbox{EeV}} \right )^{0.97} \nonumber\\
&+& \Nmu(E)\cdot S_{\mu;\rm ref}(10\:\mbox{EeV}, DX(\Xmax,\theta))
\label{eq:SparamE}
\end{eqnarray}
As the energy scaling of $\Smu$ is slightly model-dependent, we treat it as an
unknown and define $\Nmu(E)$ as the muon normalization at the energy $E$ 
with respect to the proton-QGSJetII reference at the fixed energy of 10~EeV.

\subsection{Shower fluctuations}
\label{sec:fluct}

\begin{figure}[t]
\begin{minipage}[t]{0.48\textwidth}
\center
\includegraphics[width=\textwidth]{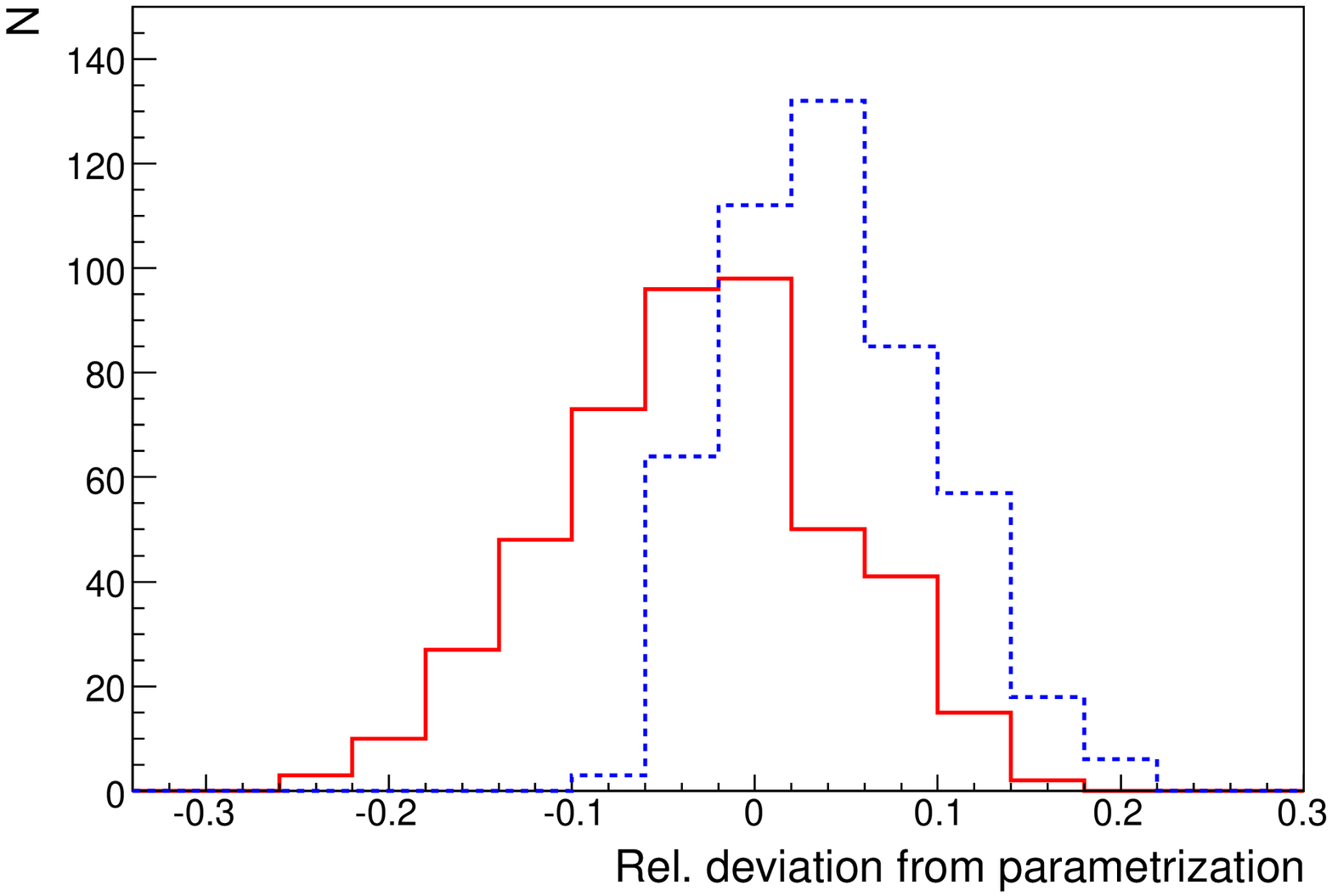}
\end{minipage}
\hfill
\begin{minipage}[t]{0.48\textwidth}
\center
\includegraphics[width=\textwidth]{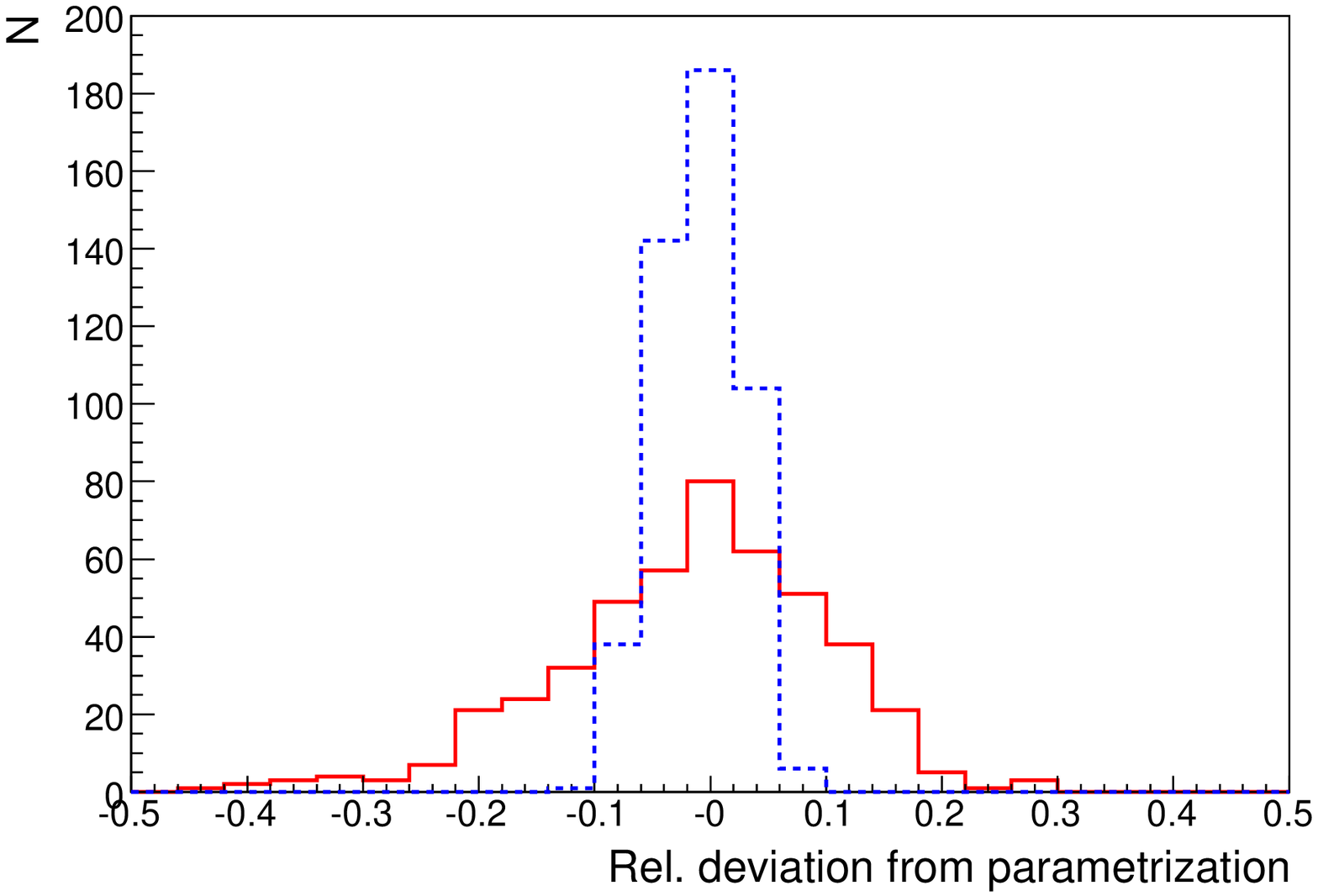}
\end{minipage}
\caption{{\it Left panel:} Distribution of the relative deviations of 
the electromagnetic ground plane signals of showers at $10^{19}$~eV 
(QGSJetII) from the parameterization (\refsec{param}). The red (solid) 
line is for proton, while the blue (dashed) is for iron. 
{\it Right panel:} The same for the muon ground plane signals.}
\label{fig:sh2sh}
\end{figure}

In addition to the overall behavior of the signals with $DX$ parameterized 
before,
both electromagnetic and muon signals show fluctuations around the
mean value. \reffig{sh2sh} shows the relative
deviations of the ground plane $\Sem$ (left panel) and $\Smu$ (right panel) 
from the parameterization for proton
and iron showers ($10^{19}$~eV, QGSJetII). These distributions contain
showers from all zenith angles; no dependence of the relative
fluctuations on zenith angle has been found. Note that the proton and
iron electromagnetic signals are slightly shifted from 0 due to the 
universality violation (\reffig{Sem}),
whereas the deviations are centered around 0 for the muon signals (we
used the corresponding muon signal normalizations for proton/iron).

The spread of the distribution shown in \reffig{Sem} and
\reffig{Smu} has a contribution from the artificial fluctuations due to the 
{\it  thinning} procedure used in the simulations.  The fluctuations due
to thinning can be estimated from vertical showers: since we expect
the same signal in all azimuth sectors, thinning
fluctuations are expected to be the dominant source of the
variance between sectors. We find $\sigma_{\rm thin}= 6.5$\% for $\Sem$ 
and 4\% for
$\Smu$. We can then subtract the uncorrelated thinning variance from the total 
signal fluctuations to obtain the shower-to-shower fluctuations.  Note that
since we compare shower signals with the parameterization of the average signal
{\it at the same distance to ground} $DX$, the
signal fluctuations shown here are not caused by the fluctuations
in the depth of shower maximum. The latter ones will induce
additional fluctuations (mainly in the electromagnetic signal) that
can be straightforwardly calculated convolving the fluctuations in $\Xmax$
with the signal parameterization.

\begin{table}[b]
\center
\begin{tabular}{c|c|c|c|c|c}
\hline
& RMS($\Sem$)$/\Sem$ & RMS($\Smu$)$/\Smu$ & $\Xmaxavg$ (10~EeV) & $\tau_X$& $\Nmu$\\
\hline
{\bf Proton} & & & & & \\
QGSJetII & 7.9\% & 11.8\% & 787.8$\:\gcmsq$ & 25.4$\:\gcmsq$ & 1 \\
Sibyll & 8.9\% & 12.4\% & 795.8$\:\gcmsq$ & 24.1$\:\gcmsq$ & 0.87 \\
\hline
{\bf Iron} & & & & & \\
QGSJetII & 5.4\% & 3.5\% & 708.7$\:\gcmsq$ & 10.9$\:\gcmsq$ & 1.40 \\
Sibyll & 4.8\% & 4.0\% & 696.5$\:\gcmsq$ & 10.2$\:\gcmsq$ & 1.27 \\
\hline
\end{tabular}
\caption{Relative shower-to-shower fluctuations of the electromagnetic 
and muon signals 
and parameters of the $\Xmax$ distribution derived from QGSJetII and Sibyll 
showers at $10^{19}$~eV. The muon signal normalization $\Nmu$ relative to proton-QGSJetII 
for the different models is also shown. Note the differences in the absolute
value of $\Xmaxavg$ and $\Nmu$, while the fluctuations are model-independent.}
\label{tab:sh2sh}
\end{table}

We also parameterized the distribution of $\Xmax$ for different primaries
and models, using the following functional form:
\begin{equation}
\frac{dN}{d\Xmax} \propto x^4\:e^{-x},\;\;0 < x < \infty;\quad 
x = \frac{\Xmax-\Xmaxavg}{\tau_X}+5,
\label{eq:dNdXmax}
\end{equation}
where $\Xmaxavg$ denotes the mean depth of shower maximum, 
and $\tau_X$ is related
to the RMS of the distribution via $\tau_X = \mbox{RMS}(\Xmax) / \sqrt{5}$.
This asymmetric distribution is found to be a good fit to the 
$\Xmax$ distributions for different primaries and models.

\reftab{sh2sh} summarizes the magnitude of fluctuations in $\Sem$ and
$\Smu$ as well as $\tau_X$ for protons and iron using different
hadronic models.  Clearly, the shower-to-shower fluctuations in signal
as well as $\Xmax$ are independent of the hadronic model considered,
but depend quite strongly on the primary particle. Hence, if measured,
fluctuations can serve as a robust, model-independent indicator of
composition. Our simulations predict that these fluctuations depend only very
weakly on energy. 

\section{Limits of universality}
\label{sec:viol}

The main discernible deviation from the universality approach adopted
here is the difference in electromagnetic signal between proton and
iron showers. This difference, which we refer to as {\it universality
  violation}, is larger than the differences found in the overall
energy deposit in the atmosphere for proton and iron showers (for which in fact
one finds the opposite effect: the so-called {\it missing energy} is
larger for iron showers, \cite{EngelAstropart}).  Since we include
muon decay products in the muon signal, this deviation is unrelated to
the differences in muon content between protons and iron.

\begin{figure}[t]
\begin{minipage}[t]{0.48\textwidth}
\center
\includegraphics[width=\textwidth]{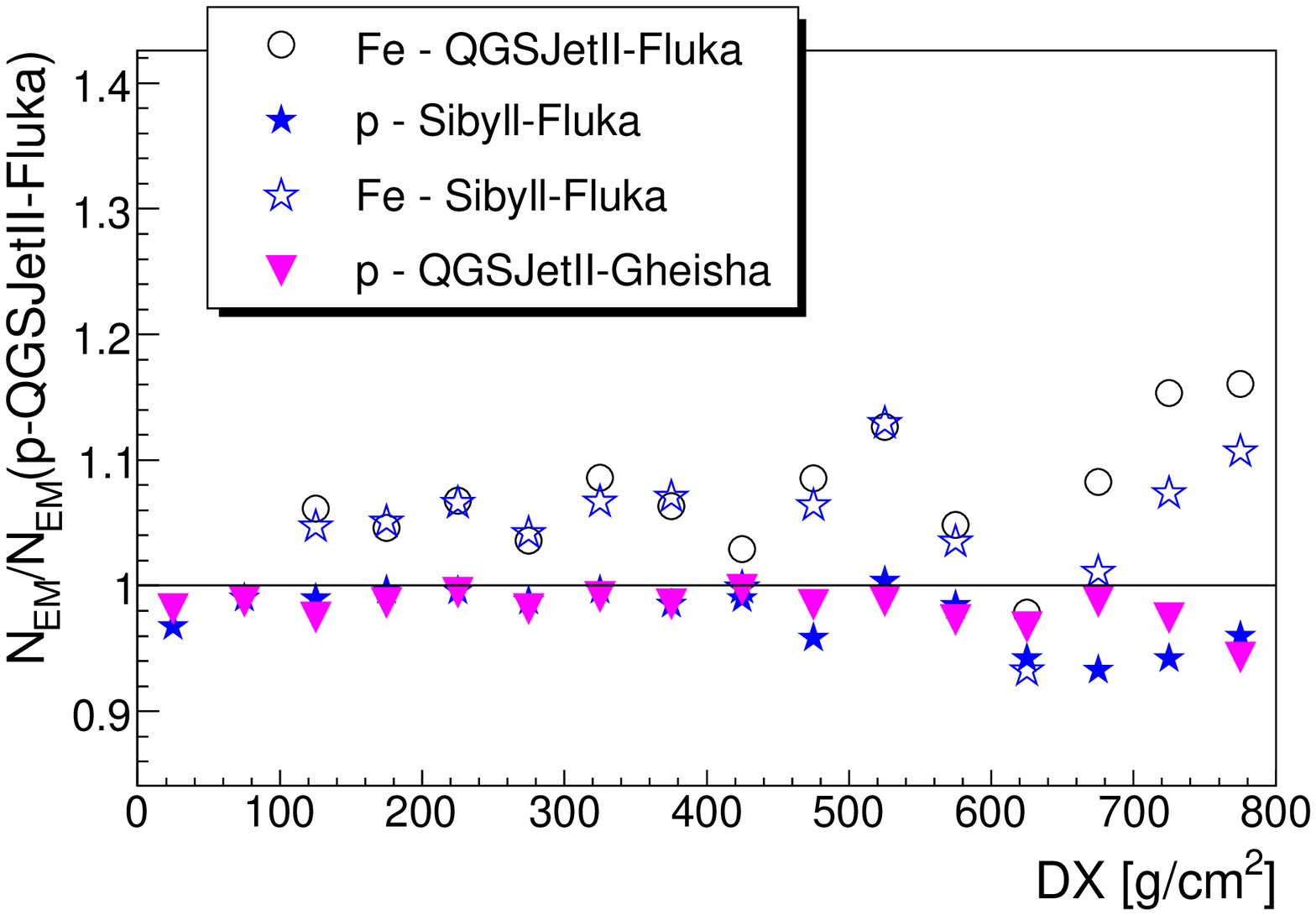}
\caption{Number flux of electromagnetic particles in the shower plane
at $r=1000$~m for different primaries and hadronic models at $10^{19}$~eV,
relative to that of proton-QGSJetII.}
\label{fig:Nemratio}
\end{minipage}
\hfill
\begin{minipage}[t]{0.48\textwidth}
\center
\includegraphics[width=\textwidth]{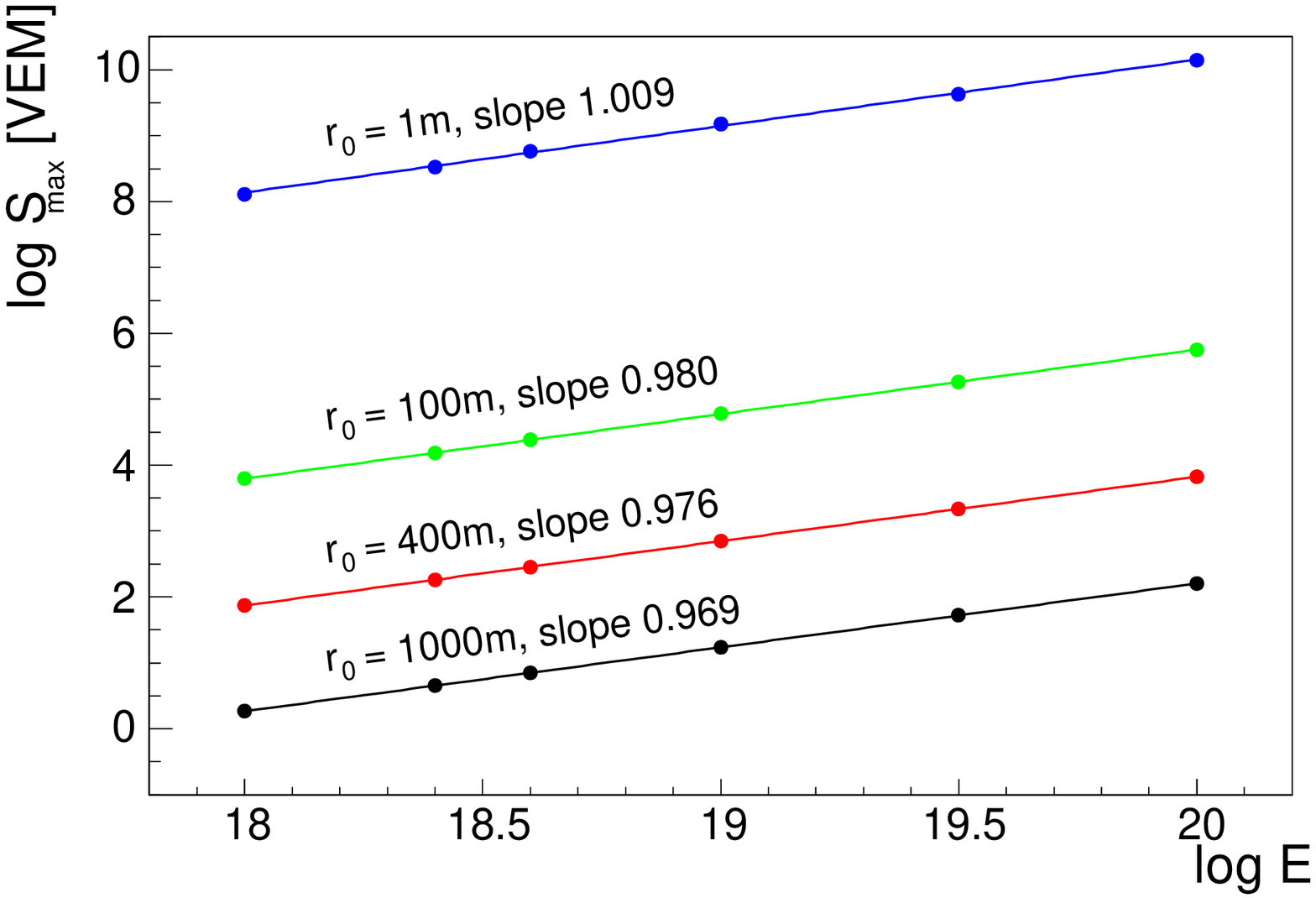}
\caption{The parameter $\Smax$ of the parameterization \refeq{gh} of 
electromagnetic shower plane signals as a function of primary energy,
for different core distances $r$. Power-law fits and the resulting exponents
are indicated.}
\label{fig:Semscaling}
\end{minipage}
\end{figure}

\reffig{Nemratio} shows the ratio of the number flux of electromagnetic
particles for different combinations of primaries/models to the
reference (proton/QGSJetII) as a function of $DX$ (again $r$=1000~m
and $E=10^{19}$~eV). The differences between protons and
iron is much smaller than in the case of the signals (\reffig{Sem}),
pointing to a slightly harder energy spectrum for electromagnetic
particles at large $r$ in iron showers compared to proton showers.
We also found that the discrepancy becomes smaller at smaller core
distances.  We have verified that the differences are independent
of the details of nuclear fragmentation of the primary iron nuclei. This means
that the nuclear binding of the 56 nucleons is not important 
for the EAS development. In other words, the superposition model holds, 
i.e., an iron shower can be considered as a superposition of 56 proton 
showers at $1/56$th of the primary energy. 

This implies that the universality
violation is due to a violation of strict linear energy scaling of the
electromagnetic signal in hadronic shower simulations:
if the electromagnetic signal scales as $E^{\alpha}$, $\alpha < 1$, then
the signal of an iron shower will be a factor of $56^{1-\alpha}$ times
larger than that of a proton shower
at the same energy. In order to explain the observed difference of $\sim15$\% 
in the shower plane signals, we would infer $\alpha\sim 0.97$.
We have parameterized the electromagnetic signals for different energies and
indeed found that the amplitude $\Smax$ of the signal (\refeq{gh}) scales as 
$E^{0.97}$ at $r=1000$~m, with $\alpha$ approaching 1 as $r\rightarrow0$ 
(\reffig{Semscaling}). Note that, by parameterizing the complete evolution
of the signal with $DX$, we take out the effects of the energy dependence of
$\Xmax$.

This violation of perfect energy scaling of the electromagnetic signal
can be due to several reasons. The injection rate of energy into the 
electromagnetic
part via $\pi^0$ decay as well as the energy spectrum of secondary $\pi^0$
might evolve with primary energy. In addition, the NKG theory of pure
electromagnetic showers also predicts a slight deviation from perfect energy
scaling of the particle flux on ground. These effects are currently under
investigation.

\section{Determining the muon normalization using the constant intensity method}
\label{sec:CIC}

One of the main challenges of a cosmic ray surface detector is to convert the
ground signal $S(r)$ to a primary energy. As mentioned in 
\refsec{param}, the universality-based signal parameterization has
three free parameters. Apart from the zenith angle $\theta$ which is
well measured along with the signal \cite{augericrcS1000}, the depth
of shower maximum $\Xmax$ and muon normalization $\Nmu$ (with respect
to the reference signal at the fixed energy of 10~EeV) remain to be
determined. Once these are known, \refeq{SparamE} provides a one-to-one
mapping of ground signal and energy, i.e. a model-independent energy
scale of the experiment.

\begin{figure}[t]
\begin{minipage}[t]{0.48\textwidth}
\center
\includegraphics[width=\textwidth]{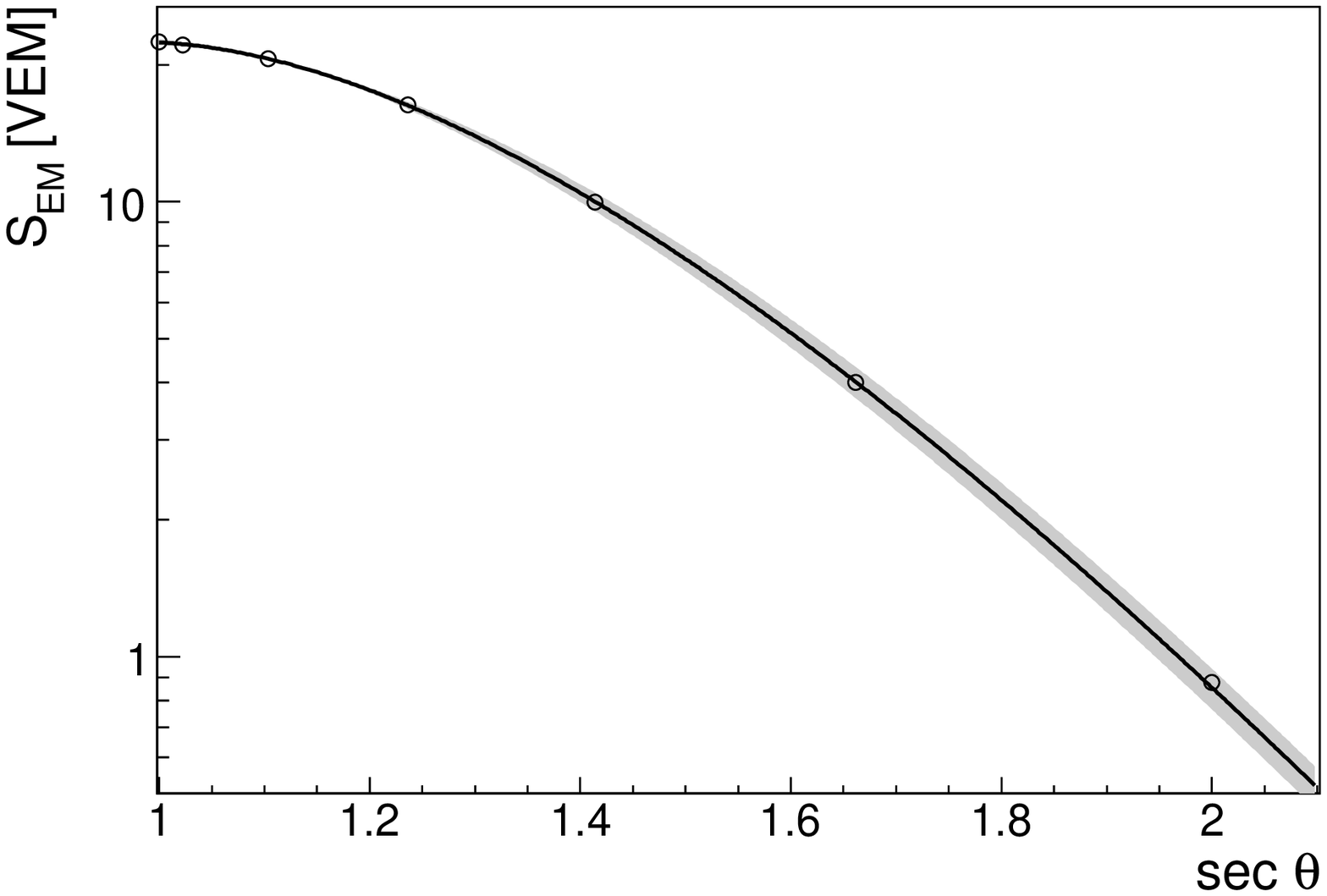}
\caption{Parameterized electromagnetic signal at $10^{19}$~eV
vs. $\sec\theta$ with $\Xmax = 750\:\gcmsq$ (black line, the dots indicate
the signal parameterized at each zenith angle). The shaded band shows the
effect on $\Sem$ of a variation of $\Xmax$ by $\pm 14\:\gcmsq$.}
\label{fig:Sem-sectheta}
\end{minipage}
\hfill
\begin{minipage}[t]{0.48\textwidth}
\center
\includegraphics[width=\textwidth]{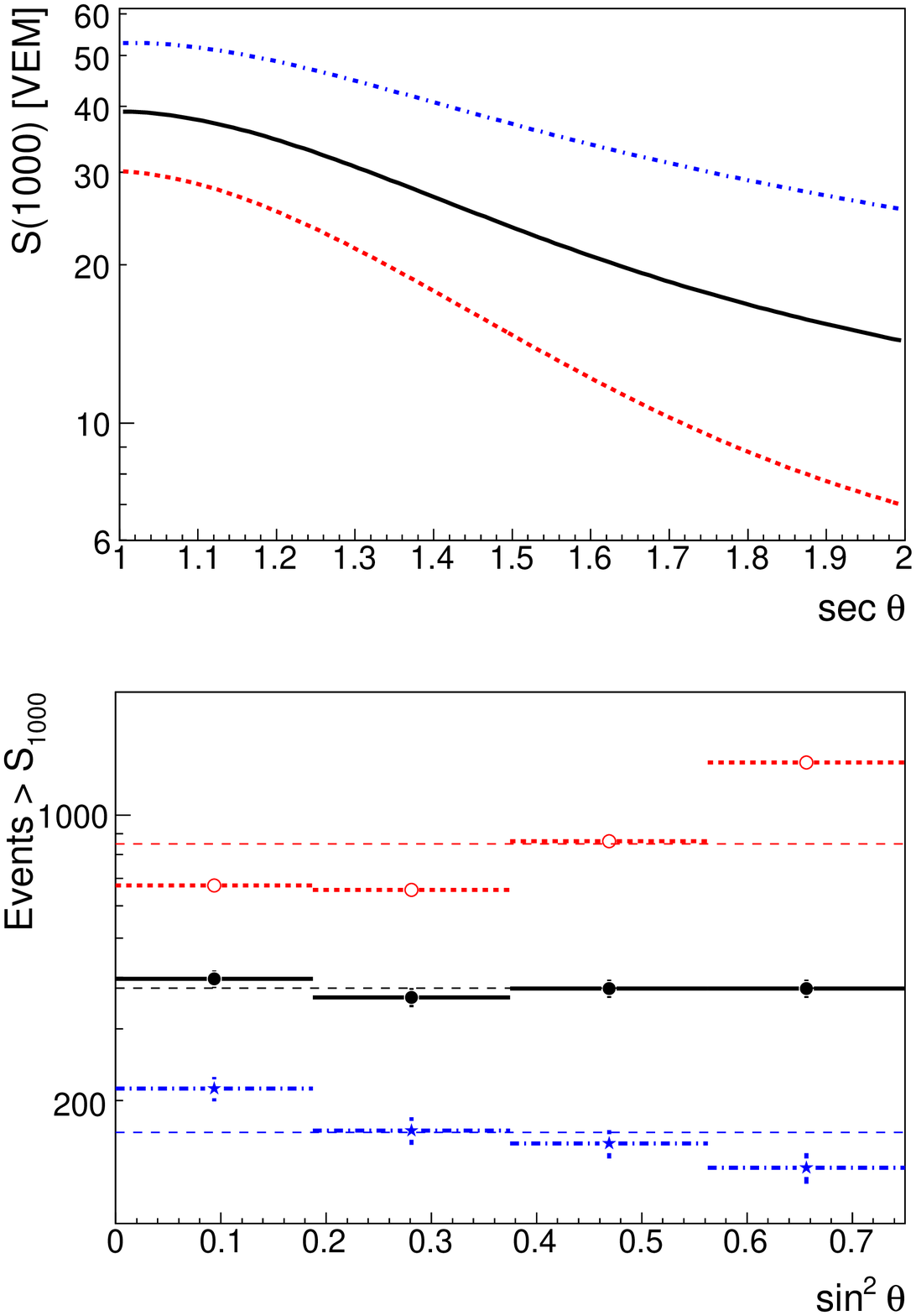}
\caption{{\it Upper panel:} the signal parameterization at 10~EeV \refeq{Sparam} 
vs. $\sec\:\theta$ for different $\Nmu$ (black/solid$-$1.1, red/dashed$-$0.5, 
blue/dash-dotted$-$2.0). {\it Lower panel:} 
histograms of number of events above the parameterized signal in equal
exposure bins, obtained from a Monte Carlo data set with a true $\Nmu$ of 1,
for the different $\Nmu$ values shown in the upper panel.}
\label{fig:CICmethod}
\end{minipage}
\end{figure}

The mean depth $\Xmaxavg$ of showers has been measured as a function of
energy from experiments using the air fluorescence technique, e.g. HiRes
\cite{hiresXmax} and Auger \cite{augericrcXmax}. The knowledge of $\Xmaxavg$ is
important as it determines the average distance to shower maximum $DX$ for a given
zenith angle, and the electromagnetic signal evolves strongly with $DX$.
The overall precision of these $\Xmaxavg$ measurements is better than 
$20\:\gcmsq$, and this small uncertainty in $\Xmaxavg$ has only a limited effect
on the estimated electromagnetic signal. \reffig{Sem-sectheta} shows the
limited effect of varying $\Xmax$ by $\pm 14\:\gcmsq$ (the current measurement 
uncertainty at 10~EeV reported by Auger \cite{augericrcXmax}) on $\Sem$ 
as a function of $\sec\theta$.

 The main uncertainty in determining the energy scale of surface detectors
 is thus in the value of $\Nmu$. Fortunately, one can make use of the
 different behavior of $\Sem$ and $\Smu$ with $DX$ (and hence,
 $\sec\theta$) to measure $\Nmu$ via the {\it constant intensity
   method} (\reffig{CICmethod}): dividing the data set into equal
 exposure bins in zenith angle, i.e., bins of $\sin^2\theta$, a
 correct signal-to-energy convertor should yield the same number of
 events in each bin with measured signal greater than the parameterized 
 signal at a fixed energy.  This is due to the isotropy 
 ($\theta$-independence) of the
 cosmic ray flux, which requires that the number of events $N(>E)$
 above a fixed energy $E$ should be equal in equal exposure bins.

\reffig{CICmethod} (upper panel) shows the zenith angle dependence of
the signal (\refeq{Sparam}) for a fixed energy of 10$^{19}$ eV and
different values of $N_\mu$. Apart from the overall change in signal,
it is evident that the smaller the
$N_\mu$, the steeper the $\theta$ dependence is. We now divide a
simulated ground detector data set with a ``true'' $\Nmu(10^{19}\rm
eV) =1$ (see \refsec{toyMC} for details on the simulation) in equal
exposure bins in zenith angle.  Given a muon normalization, we
calculate the number of events in each bin that are above a given reference
energy (here $E_{\rm ref}$=10$^{19}$~eV), according to \refeq{Sparam} with
the given $\Nmu$. We
then adjust $\Nmu(E_{\rm ref})$ in the signal parameterization
\refeq{Sparam} to the value which gives an equal number of events
$N(>S(E_{\rm ref}, \theta))$ in each zenith angle bin (lower panel in
\reffig{CICmethod}). Clearly, a too low value of $\Nmu$ results in an
excess of events at high $\theta$ (the parameterized signal  has a too
steep attenuation with $\sec\theta$), whereas a too high $\Nmu$
results in a deficit of high zenith angle events ($\sec\theta$
attenuation too shallow). In this calculation we used the $\Xmaxavg$ at 10~EeV
reported in \cite{augericrcXmax} to calculate the ground plane signals.
Note that an $\Nmu$ of 1.1 gives a flat distribution, whereas the ``true''
$\Nmu$ used is 1.0. This bias in the $\Nmu$ measurement will be addressed
in the next section.

For a range of $\Nmu$ values, we then calculate the $\chi^2$/dof of the event
histogram relative to a flat distribution in $\sin^2\theta$. Fitting a 
parabola to the function $\chi^2(\Nmu)$ yields
the best-fit $\Nmufit$ and its error $\sigma_{\Nmu}$:
\begin{equation}
\chi^2(\Nmu) = \chi^2_{\rm min} + \left ( \frac{\Nmu-\Nmufit}{\sigmaNmu} \right)^2
\label{eq:chi2}
\end{equation}

For a data set comparable to current Auger statistics
($\sim$11~000 events above 3~EeV \cite{augericrcSpectrum}), 
we expect a statistical error of $\sigmaNmu
= 0.1$. Once $\Nmu$ is known, the knowledge of $\Sem$ (within the
uncertainty of $\sim \pm 6$\% due to universality violation) determines a
model-independent energy scale, with a statistical error of
$\sigmaNmu\cdot S_{\mu;\rm ref}$ around 4\%. The constant intensity
method can be extended to other energies, using the energy-dependent
parameterization \refeq{SparamE} in \refsec{param}. This yields a measurement of 
$\Nmu(E)$, comparable to the measurement of $\Xmaxavg$ in its sensitivity 
to the primary composition. 
\reftab{errors} contains a summary of the expected
statistical and systematic errors from current and upcoming
experiments. 

\begin{figure}[t]
\centering
\includegraphics[width=0.6\textwidth]{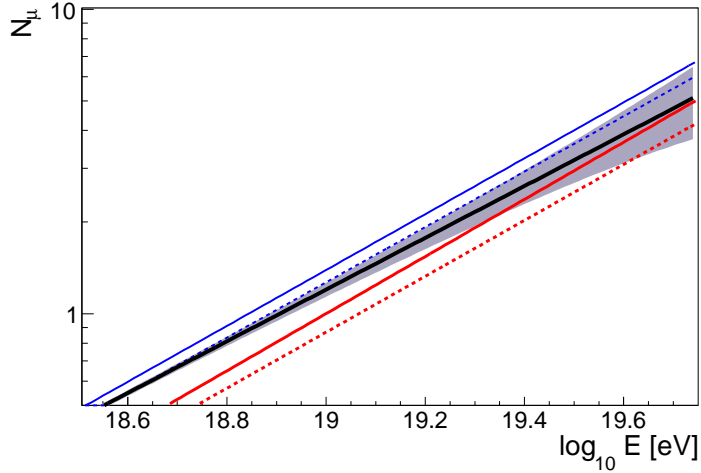}
\caption{The measured muon normalization $\Nmu$ as a function of energy 
(thick black line) with statitistical error band expected from a three-year 
Auger exposure (shaded), for a fiducial $\Nmu(E) = 1.2\: (E/10\:\mbox{EeV})^{0.85}$. Also shown are model predictions for iron 
(upper two lines, blue; 
solid$-$QGSJetII, dashed$-$Sibyll) and proton showers (lower two lines, red; 
solid$-$QGSJetII, dashed$-$Sibyll).}
\label{fig:Nmu-elrate}
\end{figure}

In \reffig{Nmu-elrate} we show possible results of this
measurement, the integral measurement of $\Nmu(E)$ (solid black line, 
corrected for the bias, see \refsec{toyMC}) and
with 1$\sigma$ statistical error band (shaded) after a three year Auger
exposure. Here, we took $\Nmu(E) = 1.2 (E/10\:\mbox{EeV})^{0.85}$ as fiducial
value. As the cosmic ray spectrum drops rapidly with energy ($\propto E^{-3}$),
the average energy of cosmic rays above a given energy threshold is very close
to that threshold.
Hence, for slow changes of the cosmic ray composition with energy, the $\Nmu$
value determined from the constant intensity method will 
reflect the actual average value of $\Nmu$ for cosmic rays at that energy
(this will be shown in the next section).
In case of an abruptly changing composition, the measured $\Nmu(E)$ will clearly 
show evidence for this. However, the interpretation of the integral $\Nmu$ 
measurement in terms of composition will have to rely on a modeling of the 
composition evolution in this case.
In addition, the measurement of $\Nmu(E)$ can place constraints on  
hadronic models, whose predictions are shown as lines in \reffig{Nmu-elrate}.

\begin{table}[b]
\center
\begin{tabular}{c|c|c}
\hline
& ~Muon normalization $\Nmu$~ & ~Energy scale~ \\
\hline
{\bf Statistical error} & & \\
current Auger & 0.1 & 4\% \\
\hline
{\bf Systematic errors} & & \\
$\Xmaxavg$ uncertainty (14$\:\gcmsq$ \cite{augericrcXmax}) & +0.05 / -0.07 & +0.5\% / -2\% \\
Universality violation & +0.01 / -0.04 & +3\% / -4\% \\
$\Nmu$ bias & $\lesssim$10\% & $\lesssim 5$\% \\
\hline
\end{tabular}
\caption{Expected statistical (for current Auger exposure) and systematic 
errors on the muon normalization
$\Nmu$ and energy scale $S(\theta=\theta_0,\:E=10\:\rm EeV)$ at 10~EeV.}
\label{tab:errors}
\end{table}

\section{Validating the constant intensity method}
\label{sec:toyMC}

To benchmark and validate the determination of $\Nmu$ via the
constant intensity method, we simulate realistic data sets based
on our parameterization of the ground signal (\refsec{param}) and its 
fluctuations (\refsec{fluct}). The fluctuations in signal as well as
$\Xmax$ could have an impact on the measurement of $\Nmu$, 
since only the average values are used to infer $\Nmu$ (\refsec{CIC}). 
Additionally, a mixed composition of the
 cosmic ray beam could bias the measurements. The purpose of this section
 is to quantify systematic uncertainties of the method described before.

 The calculation of a simulated data set proceeds as follows. Event
 energies are drawn from a spectrum $dN/dE \propto E^{-2.9}$ in the
 range $10^{17.8}-10^{20.2}$eV, while the zenith angle is drawn from
 an isotropic distribution ($\theta< 70\Deg$). 
 The primary particle type (proton or iron)
 is chosen at random according to a given mixture. The depth of shower
 maximum is then drawn from the distribution \refeq{dNdXmax} with the
 parameters for the given primary (we adopt the parameters from
 QGSJetII; this has no influence on our conclusions). An $\Nmu$ is
 determined according to the primary.  With $E$, $\theta$, $\Xmax$,
 and $\Nmu$ given, the ground signal can be determined via
 \refeq{Sparam} (we scale $\Sem$ with $E^{0.97}$, and $\Smu$ with
 $E^{0.9}$). The two signal components are fluctuated according to the
 primary (see \reftab{sh2sh}).  Finally, we cut events according to a
 simple trigger depending on the ground signal, and apply signal
 reconstruction uncertainties as reported by the Auger observatory
 \cite{augericrcS1000}. The main characteristics of the reconstruction 
 of $S(1000)$ are that it is unbiased at large 
 signals $S(1000) \gtrsim 10$~VEM, and that bias and variance increase quickly for
 signals below 10~VEM.

 A large set of simulated data sets showed that the error calculation
 according to \refeq{chi2} is a good estimator for the variance of the
 $\Nmu$ measurement. However, we found that the constant intensity
 method yields a systematic shift to higher $\Nmu$ values 
 of about 5--10\%. 
 This can be explained by trigger effects and fluctuations.  Due to the
 attenuation of the signal with zenith angle at a fixed energy, the
 resolution gets worse at large zenith angles. Additionally, upward
 fluctuations above the trigger threshold are more important at high
 zenith angles. These two effects, in the presence of a steep
 spectrum, produce a zenith angle-dependent enhancement of the number of events
 reconstructed above a given energy.
 This tends to {\it flatten} the constant intensity curve
 (\reffig{CICmethod}), which leads to a higher estimated $\Nmu$ value.
 The bias in $\Nmu$ is mainly determined by the experimental resolution
 and trigger effects. 
 It depends slightly on the primary composition, ranging from 4\% for pure 
 iron to 8\% for a mixed composition, due to the differing magnitudes of 
 shower-to-shower fluctuations. The unknown composition, and imperfect 
 knowledge of the experimental characteristics,
 lead to an uncertainty in the bias which should be included in the systematic
 error on $\Nmu$. We assume that this error will be smaller than the absolute
 value of the bias, hence $\lesssim$10\%.

Further systematics of $\Nmu$ are the violation of universality in the electromagnetic
ground signal ($\sim\pm 6$\%, \refsec{param}), and the uncertainty in the value of
$\Xmaxavg$, which translates into a further uncertainty in the electromagnetic
signal. \reftab{errors} gives a summary of the systematic errors derived
for $\Nmu$ and the energy scale (i.e., $S(\theta=\theta_0,\:E=10\; \rm EeV)$,
we take $\theta_0=38\Deg$, close to the median of the isotropic cosmic rays),
all evaluated at 10~EeV. We stress that since the universality violation
is smaller closer to the shower core, this method exhibits significantly
smaller systematics when applied to a surface detector measuring the signal
at $r=600$~m, instead of 1000~m  as considered here.

\begin{figure}[t]
\centering
\includegraphics[width=0.6\textwidth]{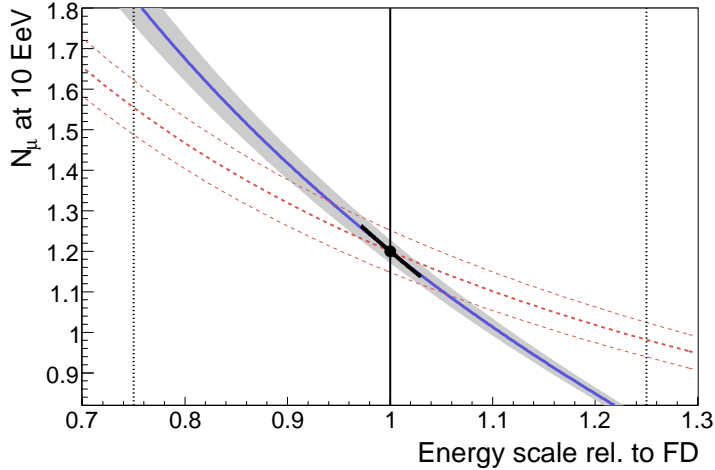}
\caption{Constraints in the $\Nmu$-energy scale plane placed by three
independent measurements: the constant intensity method (black dot with error);
vertical hybrid events (blue solid line with shaded error band); inclined
hybrid events (red dashed lines). An error of 25\% in the fluorescence 
energy scale
is indicated (vertical lines). All values are calculated for a three year
Auger-equivalent exposure. The fiducial $\Nmu$ is 1.2.}
\label{fig:Nmu-Escale}
\end{figure}

\section{Cross-checks and independent hybrid measurements}
\label{sec:hybrid}

The constant intensity method described above is
independent of the primary composition and hadronic interaction models
(within systematics), and also independent from other
energy calibrations (e.g., fluorescence telescopes).
However, it relies on a
good understanding of the electromagnetic part of air showers as well
as detector bias and resolution (\refsec{toyMC}). Hence, it is
desirable to cross-check the results of the method with independent
data.

Hybrid experiments, which simultaneously measure the ground signal as
well as fluorescence energy and $\Xmax$ on an event-by-event basis,
allow several cross-checks of the $\Nmu$ measurement. Due to the
uncertainty in the energy scale of the fluorescence detector ($\sim$
25\%), we introduce a scaling factor ($f_{FD}$) of the measured
fluorescence energy.

 Using the parameterized electromagnetic signal
 $\Sem(DX(\theta,\Xmax),\theta,E)$ (\refsec{param}), with
 $E=f_{FD} E_{FD}$ and $\Xmax$ given by the fluorescence measurement,
 we can, for a single event, determine the muon signal $\Smu$ at a fixed core 
 distance by
 subtracting the electromagnetic component from the total signal (see
 \refeq{SparamE}).  The muon normalization is then given by $\Nmu =
 \Smu/S_{\mu; \rm ref}$, where $S_{\mu; \rm ref}$ is the reference
 parameterized muon signal (proton-QGSJetII) at 10~EeV. Additionally,
 we can divide the hybrid data into two sets: vertical and inclined events, with
 zenith angles smaller and larger than 60$\Deg$, respectively. 
 In inclined events the
 electromagnetic signal at ground is essentially negligible, and the ground
 signal allows for a direct measurement of the muon signal. We can
 then calculate the mean measured $\Nmuavg$ for vertical and inclined events as
 a function of $f_{FD}$. \reffig{Nmu-Escale} shows the results for
 vertical (blue line with shaded error band) and inclined (red lines)
 events for an experiment with 3 years Auger-equivalent exposure.
 In order to measure $\Nmuavg$ with hybrid events we clearly need to constrain
 $f_{FD}$.  The black dot in \reffig{Nmu-Escale} corresponds to the
 result of the 
 constant intensity method described in the previous sections. It
 constrains $\Nmu$ {\it as well} as the energy scale. The fluorescence energy 
 scale and its current uncertainty ($f_{FD}=1.0\pm0.25$) are indicated in 
 the graph by vertical lines. 

 The crossing point of the three $\Nmu$
 measurements is an important cross-check of the universality-based method:
 only a correct description of the evolution of the electromagnetic and
 muon ground signals will lead to a unique crossing point. The value of $f_{FD}$ 
 that corresponds to the crossing point is a quite powerful 
fluorescence-independent  measurement of the energy scale. The
 statistical uncertainty of this measurement is much smaller than the current
 uncertainty on the fluorescence energy scale, and thus provides for a sensitive
 cross-check. At the same time, experimental efforts to reduce the systematic 
 fluorescence energy uncertainty are in progress \cite{Airfly}. 

Hybrid events offer several further ways to place constraints
on hadronic models. For example, since $\Smu$ and $DX$ are measured independently
for each hybrid event, the behavior of $\Smu(DX)$ can be inferred,
which contains information on the energy spectrum of muons in 
UHE air showers. In addition, the measured
fluctuations of $\Nmu$ allow for model-independent constraints
on the primary composition, if the reconstruction uncertainties are
well understood. In addition, observed anomalous $\Nmu$ values in
single events can be used to search for non-hadronic primaries. Photon
showers have a muon component of about $1/10$th of a proton shower and thus
could leave a distinctive signature in the observed $\Nmu$. However, the 
sensitivity of this method to photon showers remains to be studied
quantitatively with simulations of photon showers.

\section{Application to other experiments}
\label{sec:otherexp}

The methodology and results presented so far have been specialized to the case 
of Auger. We now discuss 
the applicability to other current and future experiments. The main 
experiment characteristic determining this application is the ratio
of the electromagnetic and muon contributions to the shower size observable
($S(1000)$ in the case of Auger). 

If the muon contribution is very
small, the signal becomes essentially independent of $\Nmu$, and only
the knowledge of $\Xmaxavg$ is needed to predict the average signal in a 
model-independent way. An experiment operating in this regime 
(such as AGASA \cite{AGASAsim}) is 
able to experimentally verify the electromagnetic signal predicted by 
simulations.

Conversely, if the muon contribution dominates
even at small zenith angles, the attenuation (i.e., $\theta$-dependence) of 
the signal becomes very small, and a model-independent separation of
the electromagnetic and muon components is impossible. The method of
determining the energy scale using the constant intensity method is then
not applicable. However, the attenuation curve can still be measured experimentally
and compared with the prediction from simulations, which depends on the 
energy spectrum of muons produced in EAS. In addition, hybrid events with an 
independent energy measurement allow for a measurement of the absolute muon
signal normalization with respect to simulations.

The ratio of muon to electromagnetic signal is determined by three factors 
in the experimental setup: {\it 1.)~The detector type:} thin scintillator
detectors have equal response to all minimum ionizing particles and thus 
operate as particle counters.  The measured number flux of particles is 
dominated by electromagnetic particles. 
Shielding can however make scintillator detectors sensitive to muons as well. 
By contrast, muons
deposit a large signal in water Cherenkov detectors. In this case, the
ratio of area to height of the water volume determines the EM$/\mu$ ratio
(flatter tanks yielding a larger ratio). {\it 2.)~The detector spacing:} the 
spacing determines the distance at which the signal is measured
\cite{Newton}. Since the lateral distribution of the muon signal is more 
spread out than the electromagnetic signal, increasing the spacing will
increase the relative muon contribution in the measured particle flux. 
{\it 3.)~The stage of shower evolution probed:} this depends on the height 
above sea level of the experiment
and the range of primary energy observed. Showers observed very far from
the shower maximum have a small electromagnetic component.

Due to different characteristics, the ground signal will be dominated by
the electromagnetic part in some experiments (e.g., AGASA, EASTop, Telescope 
Array), whereas the muon signal will contribute significantly at others
(e.g., Auger, Haverah Park). A possible quantitative criterion for the
applicability of the constant intensity method is the significance of
the signal attenuation observed (e.g., $S(\theta=0\Deg)/S(\theta=60\Deg)$)
with respect to the statistical and systematic errors in the signal
determination. This criterion corresponds to an upper limit on the relative 
muon signal contribution, which has a very weak dependence on zenith angle.

\section{Discussion and conclusion}
\label{sec:disc}

We have shown how Monte Carlo predictions of ground signals 
can be used to determine the energy scale of surface detector experiments,
indepedently of the cosmic ray composition and hadronic interaction models.
This method overcomes the otherwise unavoidable systematics of surface 
detectors due to the unknown cosmic ray composition. In addition, it allows 
for a clean measurement of the
number of muons in extensive air showers. In light of the recent detection
of a possible GZK feature in the UHECR spectrum, the energy scale of cosmic
ray experiments is of crucial importance to distinguish between different
UHECR source scenarios. Hence, it is desirable to determine the energy
scale with several methods.
The measurement of the surface detector energy scale presented here is completely 
independent of the energy scale determined from fluorescence detectors, and 
contains different systematic uncertainties.

In this paper, we explored only a single surface detector observable,
the signal $S(r)$ at a fixed distance from the shower axis. The methodology
can be extended to parameterize the signal at different distances and
azimuth angles. An extended parameterization like this can then be compared with
each detector station in a given event, increasing the number of 
observables for each event. Ideally, perhaps in combination with other
observables like the rise time \cite{risetime1,risetime2,Healy}, 
this could be used to break the 
degeneracy of $\Nmu$ and energy on an event-by-event basis for a surface 
detector alone.

It is important to note, however, that air shower universality, the basis of 
this methodology, can be violated by new mechanisms in hadronic
interactions in EAS.  Recently, the hadronic interaction model EPOS has
been introduced \cite{Pierog,PierogAIPC}.  While the predictions for
the depth of shower maximum are within the range of the previous
models considered here, EPOS shows considerable deviations in the
ground signal predictions: at 10$^{19}$ eV, the EPOS electromagnetic
signal seems to be $\sim 20$\% larger than in the other models, while
the predicted muon signal is 50--70\% higher. These differences are
due to the production of secondary baryon-antibaryon
pairs in the GeV range which is strongly increased in EPOS. These
baryons then produce more muons and a flatter lateral distribution of the signal
compared to the other models.  These predictions, while violating air
shower universality, can be constrained by observations using the
methodology presented here: by separately parametrizing the electromagnetic
and muon signals predicted by EPOS, one can infer the relative muon 
normalization with respect to EPOS which is required by the data. 
We would like to point out that EPOS
can be compared with cosmic ray data at lower energies (e.g.
KASCADE \cite{kascade}), as it was done in the past with the QGSJetII and 
Sibyll2.1 models. In addition, accelerator experiments \cite{NA61} are underway to
measure the baryon pair production at the relevant energies. 
One might hope that, once the magnitude of the
baryon-antibaryon production is understood, hadronic models will
converge to a universal prediction of the electromagnetic part as 
shown here for QGSJetII and Sibyll.

Very generally, the methodology presented here allows for a clean comparison of 
Monte Carlo simulations with air shower data, by separating shower evolution 
effects from primary composition and high-energy interactions. In applying air 
shower universality, current and future experiments  have the potential
to tightly constrain high energy hadronic models, as well as the energy scale
and mass composition of the cosmic ray beam.


\section*{Acknowledgments}
We would like to thank the members of the Pierre Auger Collaboration, in
particular Katsushi Arisaka, David Barnhill, Pierre Billoir, Jim Cronin,
Ralph Engel, Matt Healy, and Markus Risse, for
support and helpful discussions related to this work.
We are grateful to the IN2P3 computing center in Lyon, where the shower 
library used for this paper was generated.
Aaron Chou's work is supported by the U.S. Department of Energy 
under contract No. DE-AC02-07CH11359 and by the NSF under NSF-PHY-0401232. 
Lorenzo Cazon acknowledges support from
the Ministerio de Educacion y Ciencia of Spain.
This work was supported in part by the Kavli Institute for Cosmological 
Physics at the University of Chicago through grants NSF PHY-0114422 and 
NSF PHY-0551142 
and an endowment from the Kavli Foundation and its founder Fred Kavli.

\appendix

\section{Geometrical effects, ground plane and shower plane signals}
\label{sec:SP}

\begin{figure}[t]
\centering
\includegraphics[width=0.5\textwidth]{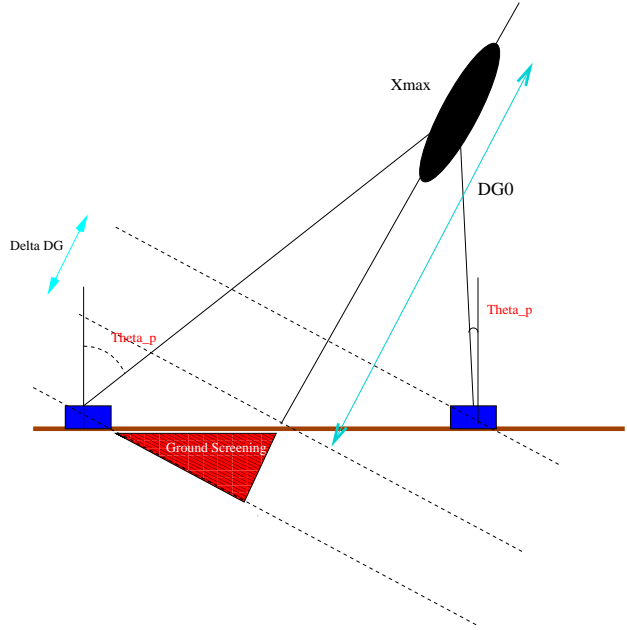}
\caption{Sketch of an inclined shower hitting the ground level. The $DX$ 
of each tank is its distance to the shower maximum, projected on the
shower axis ($DX_0$ $\pm$ $\Delta_{DX}$ in the sketch).}
\label{fig:asym_sketch}
\end{figure}

\reffig{asym_sketch} shows a sketch of an inclined shower hitting the
 ground. Two detectors at
 the same distance from the shower core correspond to different stages
 in the shower development, i.e. different $DX$ (\refeq{DX}).
 The signal size in inclined showers shows a modulation with
 $\zeta$ angle. This modulation or {\it signal asymmetry} is
 produced by a convolution of effects \cite{univicrc}. The first one is due to 
 the $\zeta$ dependence of $DX$ for inclined showers (\refeq{DX}).
 In addition, there is a geometrical effect which depends on the detector
 geometry and zenith angle of the shower.

 Consider the flux of shower particles $\v{\Phi}$ at a given $DX$ and
 $r$, so that the number of particles entering the detector is 
 $\v{\Phi}\cdot\v{A_d}$, where $\v{A_d}$ is the vector associated with the
 detector surface with $|\v{A_d}|$ equal to its area. $\v{\Phi}$
 is invariant under rotations around the shower axis, and it only
 depends on $r$ and $DX$. For vertical showers, the number of particles
 $\v{\Phi}\cdot\v{A_d}$ is invariant under rotations around the shower
 axis. For inclined showers, however, this is not the case, since $\v{\Phi}$
 is not parallel to the shower axis (\reffig{asym_sketch}).
 We therefore expect detectors at the same $r$ but different $\zeta$ angles 
 to be hit by different numbers of
 particles, even if $\v{\Phi}$ was independent of $DX$. 

Hence, the ground-plane signals depend on $DX$, $r$ and
$\theta$.  To suppress the dependence on $\theta$ and decouple shower
development effects from geometrical effects, we define the shower
plane signal as the signal generated by shower particles passing
through the top surface of a detector placed perpendicular to the
shower axis. This allows us to combine different zenith and $\zeta$
angles when plotting the shower-plane signal vs $DX$ (e.g., \reffig{Sem}). 

While the ground signals can be straightforwardly calculated from the
simulation output particles (taking into account the statistical weight $W$
due to thinning), the
shower plane signals are calculated in the following way.

Consider a particle with weight $W$ and unit momentum vector $\vphat$
hitting a sampling region at ground. We will associate a flux 
$\v{\Phi_i}$ to this particle in such a way that:
\begin{equation}
N^{GP}=W = \v{\Phi_i}\cdot \v{A_{s}^{GP} }= \Phi_i\:A_{s}^{GP}\: \vphat\cdot\vghat
\end{equation}
 where $N^{GP}$ is the number of particles hitting a ground sampling area
 $A_{s}^{GP}$ ($\sim10^5\:\rm m^2$), and $\vghat$ is the unit normal to the 
ground. Therefore,
 the flux associated with a particle of weight $W$ is given by:
\begin{equation}
\v{\Phi_i}= \frac{W}{\vphat\cdot\vghat\:A_{s}^{GP}} \vphat \\
\end{equation}

The number of particles crossing the corresponding sampling area $A_{s}^{SP}$
in the 
shower plane is given by the scalar product of this flux and a vector parallel 
to the shower axis
with a normalization equal to $A_{s}^{SP}$:
\begin{equation} 
N_i^{SP}= \v{\Phi_i}\cdot \v{A_{s}^{SP} }= W\:\frac{\vphat\cdot\vahat}{\vphat\cdot\vghat} \frac{A_{s}^{SP}}{A_{s}^{GP}} =
 W\:\frac{\vphat\cdot\vahat}{\vphat\cdot\vghat} \vahat\cdot\vghat
\end{equation}
where $\vahat$ is a unit vector along the shower axis, and we used
$A_{s}^{SP} / A_{s}^{GP} = \vahat\cdot\vghat$.

The shower-plane signal of a detector is then given by:
\begin{equation}
S^{SP}= \sum_{i} \: N_i^{SP}~\frac{A_{\rm tank}}{A_{s}^{SP}}~R(E_{p,i})
\end{equation}
where $A_{\rm tank}$ is the area of the top surface of a detector, $R(E_{p,i})$
is the detector response to a vertically incident particle with 
energy $E_{p,i}$, and the sum runs over all weighted particles falling into
the sampling area.

\bibliography{univ}
\bibliographystyle{elsart-num}

\end{document}